\documentclass[preprint,12pt]{elsarticle}

\usepackage{amssymb}

\usepackage{graphicx}
\usepackage{float}
\usepackage{mathtools}
\usepackage{braket}
\usepackage{bbm, dsfont}
\usepackage{hyperref}
\usepackage{caption}
\usepackage{subcaption}
\usepackage{cleveref}
\usepackage{xcolor}

\newtheorem{theorem}{Theorem}[section]
\newtheorem{remark}{Remark}[section]

\makeatletter
\def\ps@pprintTitle{%
	\let\@oddhead\@empty
	\let\@evenhead\@empty
	\let\@oddfoot\@empty
	\let\@evenfoot\@oddfoot
}
\makeatother

\begin{document}

\begin{frontmatter}



\title{An Exactly Solvable Model for Single-Lane Unidirectional Ant Traffic}


\author[inst1,inst2]{Ngo Phuoc Nguyen Ngoc\corref{cor1}}
\ead{ngopnguyenngoc@duytan.edu.vn}

\affiliation[inst1]{organization={Institute of Research and Development Vietnam, Duy Tan University},
            city={Da Nang},
            postcode={550000}, 
            country={Vietnam}}
\cortext[cor1]{Corresponding author.}           
\affiliation[inst2]{organization={Faculty of Natural Sciences, Duy Tan University},
	city={Da Nang},
	postcode={550000}, 
	country={Vietnam}}

\author[inst1,inst2]{Huynh Anh Thi}
\ead{huynhanhthi@dtu.edu.vn}
\author[inst3]{Nguyen Van Vinh}

\affiliation[inst3]{organization={Faculty of Information Technology, Ho Chi Minh City University of Economics and Finance},
            city={Ho Chi Minh},
            postcode={700000}, 
            country={Vietnam}}
\ead{vinhnv@uef.edu.vn}
\begin{abstract}
	In 2002, Chowdhury et al. introduced a simplified model aimed at depicting the dynamics of single-lane unidirectional ant traffic. Despite efforts, an exact solution for the stationary state of this ant-trail model remains elusive. The primary challenge arises from inherent fluctuations in the total amount of pheromones along the ant trail. These fluctuations significantly influence ant movement. Consequently, unlike conventional models involving driven interacting particles, the average velocity exhibits non-monotonic behavior with ant density. In this study, we propose an exactly solvable model for ant traffic based on the dynamics in Chowdhury et al.'s model and discuss the circumstances under which the non-monotonic trend in average velocity arises.
\end{abstract}


\end{frontmatter}


\section{Introduction}
Insects use various ways to communicate, including visual signals, sounds, and chemicals. Pheromones, in particular, play a crucial role in communication among species of insects. For example, ants use trail pheromones to guide colony members to food sources and back to the nest. The study of these insect colonies has contributed to various fields such as computer science \cite{Dorigo1999}, communication engineering \cite{Bonabeau2000}, artificial intelligence \cite{Dorigo1999-2}, micro-robotics \cite{Krieger2000}, as well as task partitioning and decentralized manufacturing \cite{Anderson1999, Anderson2000, ANDERSON2001, Ratnieks1999-2, Ratnieks1999}.

To investigate the collective movements of ants along already established paths, researchers have used different models to understand how ants move in groups. One such model, introduced by Chowdhury et al. \cite{Chowdhury2002}, as referenced in \cite{Schadschneider2011}, has been particularly useful. It helps study how ants move along paths and how they interact with each other. A generation of this model can be found in \cite{Gokce2013}. Couzin and Franks \cite{Couzin2003} developed an individual-based model, which not only accounts for the emergence of self-organized lanes but also offers insights into variations in ant flux. Moreover, a continuous model was proposed in \cite{Johnson2006}. The present study will focus specifically on Chowdhury et al.'s model, as its dynamics are interdependent with ours.

In regular traffic, vehicles decelerate as more cars fill the road, leading to congestion (see Subsection \ref{tasep}). In contrast, Chowdhury et al.'s model of unidirectional ant traffic (Subsection \ref{Chowdhury}) reveals a non-monotonic trend due to the interaction of ants' dynamics with pheromones.

The ant-trail model proposed by Chowdhury et al. \cite{Chowdhury2002} is closely related to the bus route model \cite{OLoan1998-1, OLoan1998-2} with parallel updating \cite{Chowdhury2000}. The authors of the present paper have also proposed an exactly solvable model for the bus route system \cite{Ngoc2024_2}.

\subsection{Totally asymmetric simple exclusion process}\label{tasep}
In 1970, Spitzer proposed a mathematical model for the exclusion process  inspired by biological traffic concerns \cite{MacDonald1968,MacDonald1969}. It serves as a toy model in nonequilibrium statistical physics to investigate nonequilibrium phenomena and phase transitions induced by boundaries. Recent literature on this subject is available in \cite{Chou2011, Schadschneider2011}. For a more comprehensive understanding of the process, readers can consult \cite{Katz1984, Liggett2005, MacDonald1968, Schutz2001, Schutz2019, Spitzer1970}.

The exclusion process  is a continuous-time Markov chain, wherein random particles must obey the exclusion principle, ensuring they do not occupy the same position simultaneously. Furthermore, particles can only move from one lattice site to an adjacent site if the destination site is unoccupied. On the lattice, we denote a vacant site by 0 and a site occupied by a particle by 1. This study focuses exclusively on variations of the one-dimensional Totally Asymmetric Simple Exclusion Process (TASEP), characterized by the following simple dynamics
\begin{equation}\label{TSEP}
	10 \to 01 \text{ with rate } 1.
\end{equation}

\begin{figure}[H]
	\centering
	\begin{subfigure}[b]{0.4\textwidth}
		\centering
		\includegraphics[width=\textwidth]{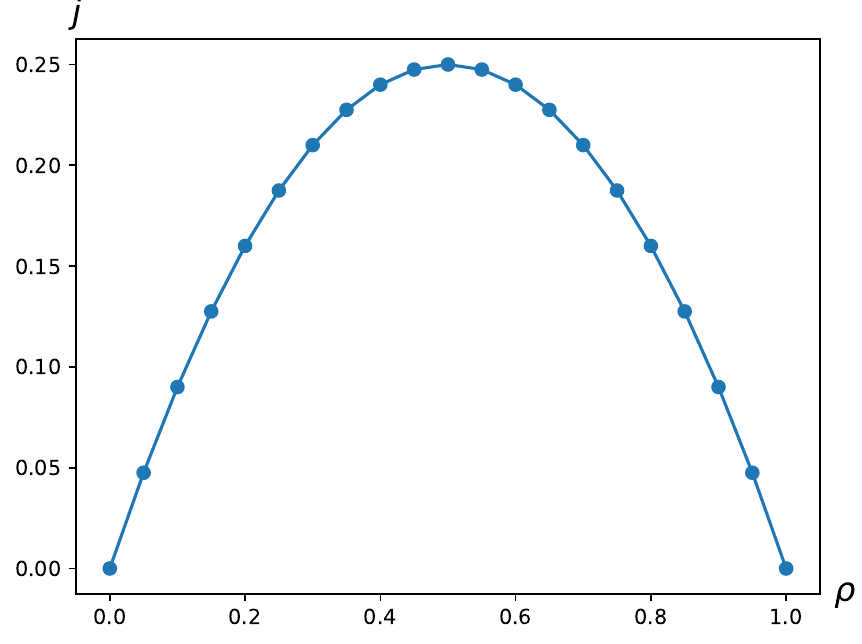}
		\caption{Average current $j$}
		\label{fig:j_seep}
	\end{subfigure}
	\begin{subfigure}[b]{0.4\textwidth}
		\centering
		\includegraphics[width=\textwidth]{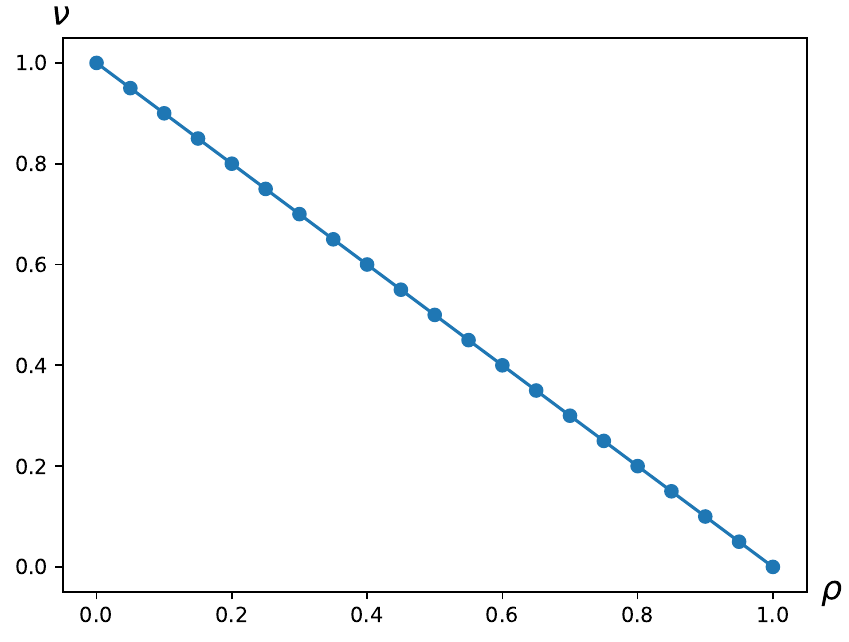}
		\caption{Average velocity $\nu$}
		\label{fig:v_seep}
	\end{subfigure}
	\caption{The average particle current $j$ and velocity $\nu$ of TASEP vary as functions of density $\rho$. The current exhibits symmetry and the velocity $\nu$ decreases linearly with density $\rho$.}
	\label{fig:jv_seep}
\end{figure}

In a lattice with periodic boundary conditions, the stationary distribution of the TASEP is uniform. In the thermodynamic limit, the stationary particle velocity $\nu$ and current $j$ are determined as functions of the particle density $\rho$, as follows
\begin{align}\label{current_d0}
	\nu = 1- \rho,\  \ j = \rho(1 - \rho).
\end{align}

Fig. \ref{fig:jv_seep} illustrates that the stationary current of particles demonstrates symmetry, while the stationary velocity follows a monotone decreasing pattern. Additionally, in the steady state particle distribution, correlations are notably absent. These characteristics contrast with findings in biological traffic models, as studied by \cite{Belitsky2019-1, Chowdhury2002, Epshtein2003-1, Epshtein2003-2}, where particle interactions on the same lattice tend to be cooperative. Furthermore, these features are also not observed in the model proposed by Chowdhury et al. \cite{Chowdhury2002} for ants movement along established trail, as discussed in the next subsection.

\subsection{Chowdhury et al.'s model}\label{Chowdhury}
\begin{figure}[ht]
	\centering
	\includegraphics[width=0.97\textwidth]{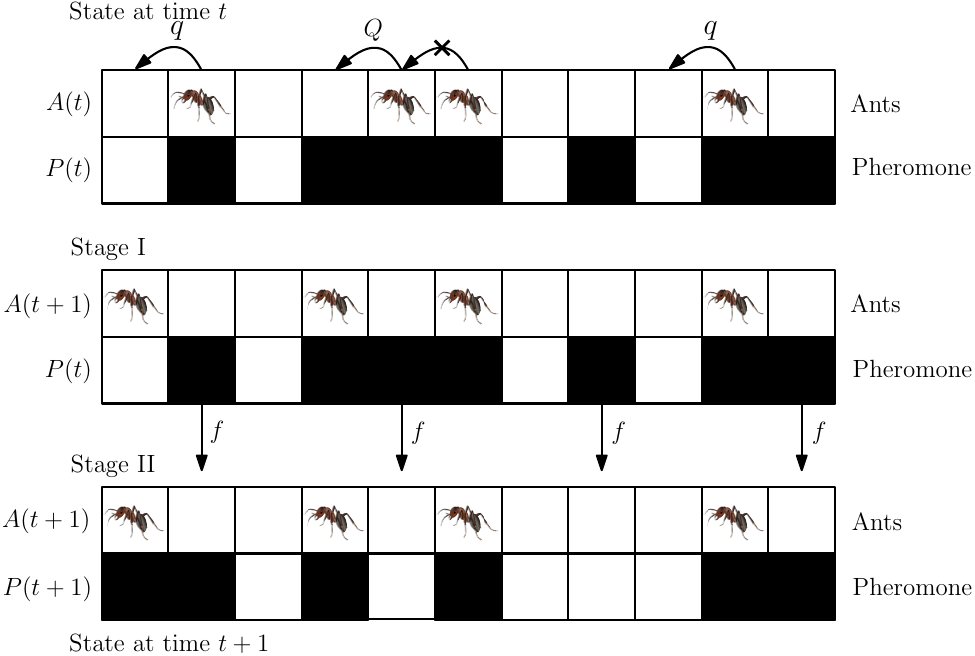}
	\caption{Illustration depicting typical configurations and the update process sketched in \cite{Chowdhury2002}. At the top of the figure, it shows a configuration at time $t$ before stage I of the update, along with the possibility of ants jumping with assigned probabilities. In the middle, the configuration after one potential realization of stage I is shown, where two ants move to the left while others remain unmoved. Additionally, in the middle of the figure, it also depicts the pheromones that may evaporate during stage II of the update scheme. At the bottom, one potential configuration after the realization of stage II is shown, indicating the evaporation of two pheromones and the creation of one pheromone due to ant motion.}
	\label{fig:model1}
\end{figure}

The unidirectional ant-traffic model (ATM) proposed by Chowdhury et al. \cite{Chowdhury2002} extends the concept of the TASEP \eqref{TSEP}. In this model, the motion of ants is coupled with a secondary field representing the presence or absence of pheromones (see Fig. \ref{fig:model1}). Specifically, the probability of ants hopping is now influenced by the detection of pheromones at the destination site, leading to a higher probability of movement. Additionally, the dynamics of the pheromones are defined: they are produced by ants, and free pheromones evaporate with a probability of $f$ per unit of time. The model comprises two stages: in Stage I, ants are allowed to move, while in Stage II, pheromones are permitted to evaporate. The authors implement parallel updates for the dynamics in each stage, meaning that stochastic dynamic rules are simultaneously applied for all ants and pheromones, independently.

In the one-dimensional ATM, each site represents a cell capable of hosting one ant at most. These lattice sites are indexed as $i$ ($i = 1, 2, ..., L$), with $L$ denoting the lattice's length. At each site $i$, one associates two binary variables, $A_i$ and $P_i$. The variable $A_i$ takes a value of 0 or 1, indicating whether the cell is vacant or occupied by an ant, respectively. Similarly, $P_i = 1$ signifies the presence of pheromone in cell $i$, while $P_i = 0$ indicates its absence. Thus, the model encompasses two sets of dynamic variables: $A(t) \equiv (A_1(t), A_2(t),..., A_i(t), ..., A_L(t))$ and $P(t) \equiv (P_1(t), P_2(t),..., P_i(t),..., P_L(t))$. The complete state, or configuration, of the system at any given time $t$ is fully described by the combined set $\{A(t), P(t)\}$. Before restating the dynamics of Chowdhury et al.'s model, for later convenience, we assume ants only move leftward.

\vspace{0.3cm}

\noindent \textbf{Stage I: Motion of ants.} In this stage, all ants are allowed to move simultaneously following the assigned probabilities. When there are no neighboring ants to the left, the probability of an ant's forward movement (to the left as previously assumed) is either $q$ or $Q$, depending on whether the target site contains pheromones. Here, $q$ represents the average speed of an ant without pheromones directly ahead of the ant, while $Q$ represents the speed with pheromones at the neighboring site to the left. Considering the configuration of the system $\{A,P\}$, let us assume an ant is located at site $i$, represented as $A_i(t) = 1$. The rates of the ant's movement to site $i-1$ can be expressed as
\begin{equation}
	\begin{cases}
		01 \to 10 \text{ with rate } q \text{ if } P_{i-1}(t) = 0, \\
		01 \to 10 \text{ with rate } Q \text{ if } P_{i-1}(t) = 1.
	\end{cases}
\end{equation}

\noindent\textbf{Stage II: Evaporation of pheromones.} This stage follows Stage I. Cells where ants have left retain pheromones. Additionally, pheromones on sites without ants are free to evaporate. This evaporation is considered a random process happening at an average rate of $f$ per unit of time. Thus, given at time $t$, there is pheromone at site $i$, i.e., $P_i(t) = 1$, one has

\begin{equation}
	\begin{cases}
		P_i(t+1) = 0 \text{ with rate } f,\\
		P_i(t+1) = 1 \text{ with rate } 1- f.
	\end{cases}
\end{equation}

As noted earlier, ant movement incorporates the presence of trail pheromones, leading to non-monotonic behavior in the average velocity of ants for small values of $f$. This phenomenon is illustrated in Figs. 2 and 3 of \cite{Chowdhury2002}. This contrasts with the behavior observed in the TASEP \eqref{TSEP} and other vehicular traffic models, where inter-vehicle interactions impede motion, resulting in a monotonic decrease in average vehicle speed with increasing density. The non-monotonic behavior in the average velocity remains true even when parallel updating is replaced by random-sequential updating \cite{Nishinari2003}. 

\begin{figure}
	\centering
	\begin{subfigure}[b]{0.93\textwidth}
		\centering
		\includegraphics[width=\textwidth]{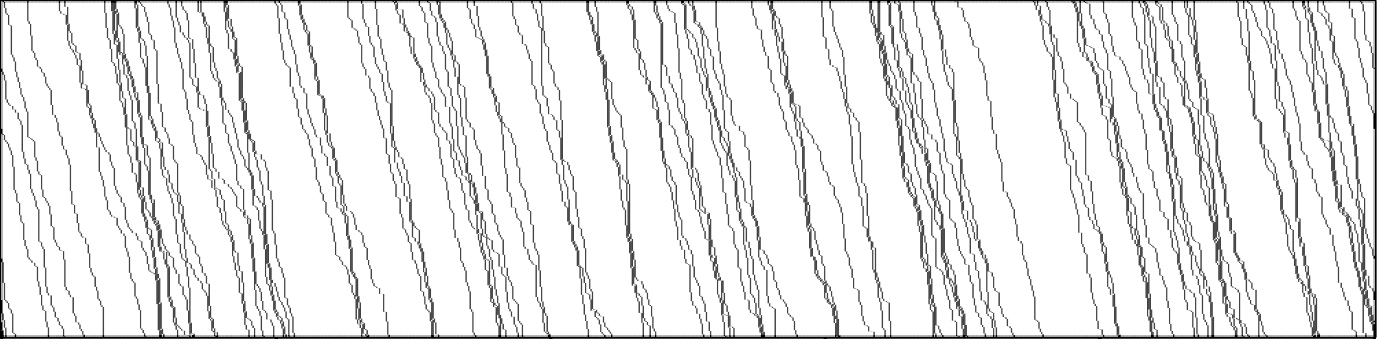}
		\caption{$f = 0.5$}
		\label{time_space_plot_a}
	\end{subfigure}
	\hspace{0.3cm}
	\begin{subfigure}[b]{0.93\textwidth}
		\centering
		\includegraphics[width=\textwidth]{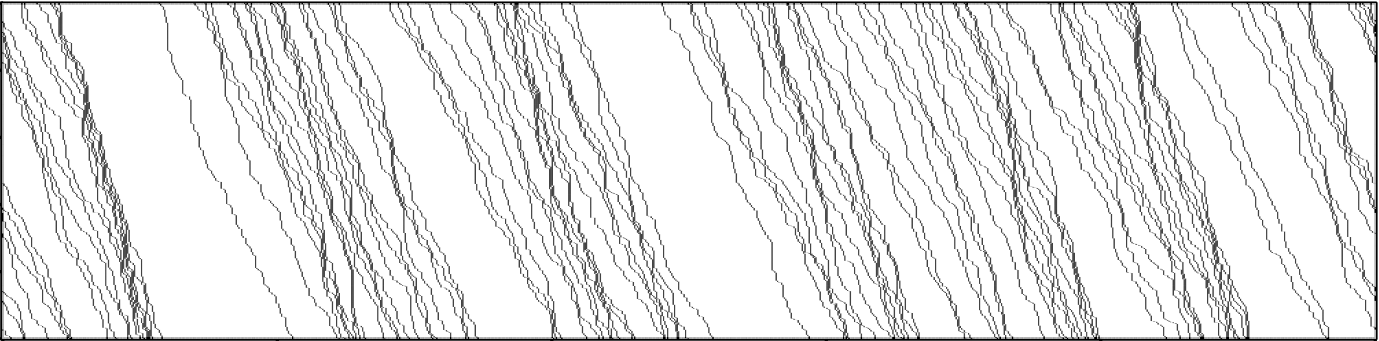}
		\caption{$f = 0.0005$}
		\label{time_space_plot_b}
	\end{subfigure}
	\caption{Space-time plots show that loose clusters do not form in the model by Chowdhury et al. with random-sequential updating. The figure depicts the state of a system with $L = 1000$ cells and $N = 100$ ants at later times, with translocation rates of $Q = 1$ and $q = 0.5$. The evaporation probability of the pheromones is $f = 0.5$ in the top panel and $f = 0.0005$ in the bottom panel.}
	\label{time_space_plot}
\end{figure}

Another interesting feature of Chowdhury et al.'s model is the loose cluster property in the low-density regime, see Fig. 5 in \cite{Schadschneider2005}. The term "loose" means there are small gaps between successive ants in the cluster, making it appear like a typical compact cluster from a distance. However, this property does not hold true when the model uses random-sequential updating, see Fig. \ref{time_space_plot} for the time-space plot. Additional sources supporting the model can be found in \cite{John2008, John2009}, and \cite{Schadschneider2011}.

%

\subsection{Motivations and Organization}
As noted in \cite{Belitsky2019-1,Schadschneider2011}, an exact solution for Chowdhury et al.'s model remains elusive. In response to this challenge, two approximate theories have been proposed. The first utilizes a mean-field approach \cite{Chowdhury2002,Nishinari2003}. However, this method encounters limitations in the intermediate density regime, where dynamics are governed by loose clusters that cannot be accurately represented by mean-field theories assuming a uniform distribution of ants. The second approach involves mapping the system onto a zero-range process (ZRP) \cite{Kunwar2004,Nishinari2003}, as the model's dynamics closely resemble those of a ZRP. This mapping, akin to the procedure in TASEP, identifies particles (ants) with the sites of the ZRP, while the gaps between the ants correspond to the particles of the ZRP \cite{Evans2005, Ngoc2024}.

The challenge lies in performing an exact analysis of ATM due to fluctuations in the total amount of pheromones on the ant-trail. Fortunately, a model proposed by Belitsky and Schütz for the RNA Polymerase \cite{Belitsky2019-1, Belitsky2019-2} explains the collective pushing phenomenon of traffic in biology and successfully captures fluctuations of this kind. In this work, we modify the model to suit our specific setting.

The remainder of the paper is structured as follows: Section \ref{model} introduces the dynamics and stationary distribution of the model, presenting the main result, Theorem \ref{main_theorem}. The proof of this result is detailed in Appendix \ref{condition_of_model}. In Section \ref{current_velocity}, we analyze the average current and velocity of ants. Following this analysis, Section \ref{discussion} delves into the conditions under which the non-monotonic trend in the average velocity of ants arises. Finally, we give conclusions in Section \ref{conclusion}.

\section{Ant-trail model with random-sequential updating}\label{model}
As the same in Chowdhury et al. model \cite{Chowdhury2002}, instead of exploring the emergence of ant trails, our focus is on observing ant traffic along an already established trail. Unlike the model, which employs parallel dynamics, our model utilizes a random-sequential update rule. 

Our approach is the following. Initially, we propose dynamics of the ATM on a one-dimensional envisioned  lattice with a length of $L$, where individual lattice sites are sequentially numbered from 1 to $L$. We assume that ants cannot overlap, and there exists an interaction between these ants, akin to the behavior observed in the TASEP for particles. However, our goal is to establish a more complex interaction rule, leading to emergent behaviors beyond what the TASEP can demonstrate. These interactions are defined by explicit formulas for the transition rates. We propose that the stationary distribution of our model adheres to a specific form. Finally, we determine the constraints on parameters in the transition rates to ensure that the envisioned distribution indeed represents the stationary distribution for the dynamics specified by those rates.

\subsection{The model state space and dynamics}
\textbf{Types of particles:} We begin this subsection by introducing the types of particles. In our context, cells on the ant trail that are occupied by ants are considered as empty sites. Thus, a site occupied by an ant is designated as state 0, representing an empty location. A site without any ant but containing pheromone is denoted as state 2, while a completely unoccupied site (totally empty) is labeled as state 1. In this setting, one might consider that a particle-hole transformation has been performed. However, in this model, cells containing pheromone are not entirely considered as holes, unlike in TASEP.

It is important to emphasize that in this context, ant motion aligns with the movement of empty sites. Moreover, each jump made by particle 0 (ant) corresponds to a jump made by particle 1 (empty cell) or 2 (cell with pheromone but without the presence of an ant). Therefore, to analyze ant movement, one can study the motion of particle types 1 and 2.

In typical behavior, we assume that particles of both types 1 and 2 move to the right, while empty sites (representing ants) shift to the left. With these particle type notations in mind, the dynamics of ants with random-sequential updating in Chowdhury et al.'s model during Stage I, and the evaporation of pheromones during Stage II in the previous section, can now be simply expressed as follows:

\begin{equation}\label{Chowdhury_rates}
	\begin{cases}
		10 \to 02,\ \ &\text{with rate } q,\\
		20 \to 02,\ \ &\text{with rate } Q,\\
		2 \to 1,\ \ &\text{with rate } f.
	\end{cases}
\end{equation}

Before presenting the dynamics of our model, note that \eqref{Chowdhury_rates} pertains to Chowdhury et al.'s model, which employs random-sequential updating rules.\\

\textbf{Ant interactions:}  Before delving into the dynamics of ants and pheromones, let us establish the state space of our model. Given our focus on the collective movement of ants, we will assume periodic boundary conditions, where the lattice $\mathbb{T}_L$ comprises $L$ sites, with $i+L \equiv i \mod L$. Considering that there are $L-N$ ants on this lattice, where $N$ represents the total number of particles of types 1 and 2, we represent a system configuration as $\boldsymbol{\eta} = (\eta_1, \ldots, \eta_L)$, where $\eta_j \in \{0,1,2\}$ for $j=1,2,\ldots,L$, signifying an allowed arrangement of particles.

In this paper, we assume that the transition rates of ants depend on the surrounding environment, specifically on the presence of pheromones and other ants. Thus, in our context, the transition rates of a particle are contingent upon the current positions of ants and pheromones. In other words, the transition rates in our model, corresponding to the rates $q$, $Q$, and $f$ in Chowdhury et al.'s model as given in \eqref{Chowdhury_rates}, rely on the current configuration.

\begin{figure}[ht]
	\centering
	\includegraphics[width=0.37\textwidth]{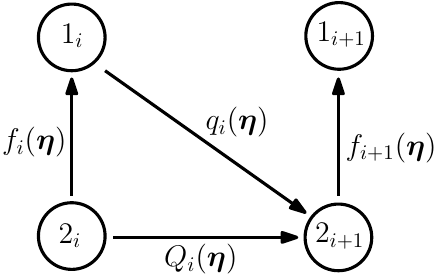}
	\caption{Minimal reaction scheme of particle translocation. A particle can move from the site $i$ to site $i+1$ provided it is in state 1 (empty cell) and change its state to 2 (cell with pheromone but without any ant located on it) with an effect rate $q_i(\boldsymbol{\eta})$. Similarly, if the particle is in state 2 at site $i$, it can move to site $i+1$ with a rate of $Q_i(\boldsymbol{\eta})$, maintaining its state. Additionally, at site $i$, a particle in state 2 can change to state 1 without moving indicating pheromone fade away, and this transition occurs with a rate of $f_i(\boldsymbol{\eta})$.}
	\label{single_ant_dynamics}
\end{figure}

Given a configuration $\boldsymbol{\eta}$, the transition rates of each particle can vary since each rate depends on its specific surrounding environment. To emphasize this dependency, we denote the transition rates of the $i$-th particle as $q_i(\boldsymbol{\eta})$, $Q_i(\boldsymbol{\eta})$, and $f_i(\boldsymbol{\eta})$. Refer to Fig. \ref{single_ant_dynamics} for an illustration of the dynamics involving a single particle of both types.

Note that an allowed configuration $\boldsymbol{\eta} = (\eta_1,...,\eta_L)$ can be defined using both a position vector $\textbf{x} = (x_1,...,x_N)$ and a state vector $\textbf{s} = (s_1,...,s_N)$ for the particles. Here, $x_i$ represents the coordinate of the $i$-th particle on the lattice, which is calculated modulo $L$, while $s_i$ represents the state of this particle, belonging to $\{1,2\}$. It is important to note that $x_i$ denotes the position of particle $i$,  rather than representing the position of an ant. 

The advantage of employing the position vector lies in its ability to express the distance between two particles using the Kronecker-$\delta$ function, defined as follows:
\begin{equation}
	\delta_{u,v} =
	\begin{cases}
		1, & \text{if } u = v,\\
		0, & \text{if } u \neq v,
	\end{cases}
\end{equation}
for any $u, v$ belonging to any set. For example, to indicate that the $i$-th particle at position $x_i$ and the $(i+1)$-th particle at position $x_{i+1}$ are located at neighboring sites, we write $\delta_{x_{i+1}-x_i,0}$. The two particles are indeed located at neighboring sites if $\delta_{x_{i+1}-x_i,0} = 1$; otherwise, it is 0.

\begin{figure}[h]
	\centering
	\begin{subfigure}[b]{0.27\textwidth}
		\centering
		\includegraphics[width=\textwidth]{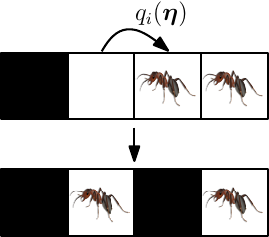}
		\caption{$q_i(\boldsymbol{\eta}) = q^\star$}
		\label{jump_rate_q_a}
	\end{subfigure}
	\hspace{0.9cm}
	\begin{subfigure}[b]{0.27\textwidth}
		\centering
		\includegraphics[width=\textwidth]{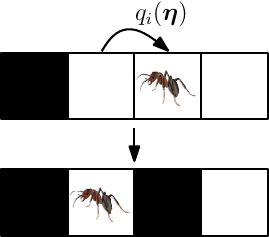}
		\caption{$q_i(\boldsymbol{\eta}) = q^\star(1 + q^{\star01})$}
		\label{jump_rate_q_b}
	\end{subfigure}
	\caption{Here are a few illustrations of the translocation rate $q_i(\boldsymbol{\eta})$ for a particle in state 1. In these visual representations, empty cells indicate that the particles are in state 1, while filled cells represent state 2. Cells with ants are considered empty sites.}
	\label{jump_rate_q}
\end{figure}
\begin{figure}[h]
	\centering
	\begin{subfigure}[b]{0.27\textwidth}
		\centering
		\includegraphics[width=\textwidth]{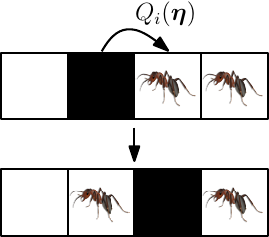}
		\caption{$Q_i(\boldsymbol{\eta}) = Q^\star$}
		\label{jump_rate_Q_a}
	\end{subfigure}
	\hspace{0.9cm}
	\begin{subfigure}[b]{0.27\textwidth}
		\centering
		\includegraphics[width=\textwidth]{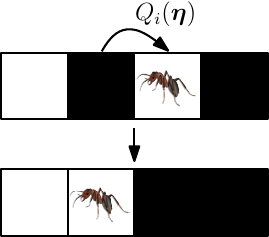}
		\caption{$Q_i(\boldsymbol{\eta}) = Q^\star(1 + Q^{\star01})$}
		\label{jump_rate_Q_b}
	\end{subfigure}
	\caption{Here are a few illustrations of the translocation rate $Q_i(\boldsymbol{\eta})$ for a particle in state 1. In these visual representations, empty cells indicate that the particles are in state 1, while filled cells represent state 2. Cells with ants are considered empty sites.}
	\label{jump_rate_Q}
\end{figure}
\begin{figure}[H]
	\centering
	\begin{subfigure}[b]{0.3\textwidth}
		\centering
		\includegraphics[width=\textwidth]{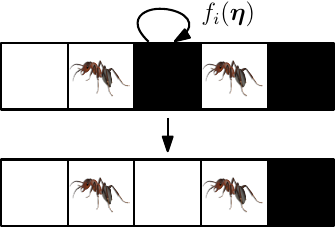}
		\caption{$f_i(\boldsymbol{\eta}) = f^\star(1+f^{10\star} + f^{\star01})$}
		\label{jump_rate_f_a}
	\end{subfigure}
	\hspace{0.9cm}
	\begin{subfigure}[b]{0.3\textwidth}
		\centering
		\includegraphics[width=\textwidth]{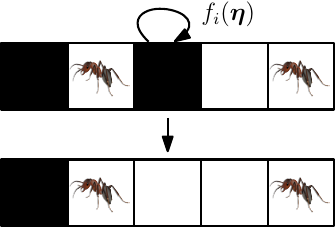}
		\caption{$f_i(\boldsymbol{\eta}) = f^\star(1 + f^{10\star}+f^{\star1})$}
		\label{jump_rate_f_b}
	\end{subfigure}
	\caption{Here are a few illustrations of the translocation rate $f_i(\boldsymbol{\eta})$ for a particle in state 1. In these visual representations, empty cells indicate that the particles are in state 1, while filled cells represent state 2. Cells with ants are considered empty sites.}
	\label{transition_rate_f}
\end{figure}

Now we are in a position to state the rates influenced by the surrounding environment. Consider the $i$-th particle, which undergoes a rightward jump with rates $q_i(\boldsymbol{\eta})$ and $Q_i(\boldsymbol{\eta})$, depending on whether its state is 1 or 2, respectively. Moreover, if its state is 2, there exists an additional possibility for transitioning to state 1 due to evaporating pheromone, governed by the rate $f_i(\boldsymbol{\eta})$. The rates of the models described above are given by
\begin{align}
	q_{i}(\boldsymbol{\eta}) & =\delta_{s_{i},1}q^{\star}(1+\delta_{x_{i}+2,x_{i+1}}q^{\star01})(1-\delta_{x_{i}+1,x_{i+1}}),\label{rate_1}\\
	Q_{i}(\boldsymbol{\eta}) & =\delta_{s_{i},2}Q^{\star}(1+\delta_{x_{i}+2,x_{i+1}}Q^{\star01})(1-\delta_{x_{i}+1,x_{i+1}}),\label{rate_2}\\
	f_{i}(\boldsymbol{\eta}) & =\delta_{s_{i},2}f^{\star}(1+\delta_{x_{i-1},x_{i}-1}f^{1\star}+\delta_{x_{i}+1,x_{i+1}}f^{\star1}+\delta_{x_{i-1},x_{i}-2}f^{10\star}+\delta_{x_{i}+2,x_{i+1}}f^{\star01}).
	\label{rate_3}
\end{align}
More precisely, concerning the description of the rates \eqref{rate_1}--\eqref{rate_3}, $\delta_{s_i,\alpha}$, with $\alpha \in \{1,2\}$, indicates the state of a particle. The parameters $q^\star$, $Q^\star$, and $f^\star$ represent the transition rates for free particles, playing roles similarly to the rates $q, Q, f$ in Chowdhury et al.'s model in \eqref{Chowdhury_rates}, respectively. To explain the roles of these parameters, take $q^\star$ as an example.  If a particle is in state 1 and the next two sites in front of it are free (i.e., these two sites are occupied by two ants), then the jump rate $q_i(\boldsymbol{\eta})$ of the particle to the right is $q^\star$ (see Fig. \ref{jump_rate_q_a}).  In equations \eqref{rate_1} and \eqref{rate_2}, the term $(1 - \delta_{x_{i}+1,x_{i+1}})$ indicates that the effectiveness of the jump rates $q_{i}(\boldsymbol{\eta})$ and $Q_{i}(\boldsymbol{\eta})$ depends on whether the next right site of particle $i$ is unoccupied. This term is equal to 0 when the $i$-th and $(i+1)$-th particles are located at neighboring sites. Regarding the superscripts, the symbol $\star$ indicates the position of particle $i$, alongside the symbol 0 representing an unoccupied site and the symbol 1 representing an occupied site. Thus, $q^{\star01}$ and $Q^{\star01}$ represent the contribution to $q_{i}(\boldsymbol{\eta})$ and $Q_{i}(\boldsymbol{\eta})$ when the $i$-th particle is positioned one empty site away from particle $(i+1)$.  The condition of a distance of 1 between the two particles is specified by the term $\delta_{x_i +2, x_{i+1}}$, which equals to 1 only when the two particles are separated by a distance of 1. Refer to Figs. \ref{jump_rate_q} and \ref{jump_rate_Q} for examples of the rates $q_i(\boldsymbol{\eta})$ and $Q_i(\boldsymbol{\eta})$. The role of parameters $q^{\star01}$ and $Q^{\star01}$ is discussed below. Similarly, the contributions of parameters $f^{1\star}$, $f^{\star1}$, $f^{10\star}$, and $f^{\star01}$ to the rate $f_i(\boldsymbol{\eta})$ can be explained. For example, the value $f^{\star1}$ is added to the rate when there is a particle at site $x_i +1$, as indicated by the term $\delta_{x_{i}+1,x_{i+1}}$. See Fig. \ref{transition_rate_f} for some examples of the rate $f_i(\boldsymbol{\eta})$.

\textbf{Neighboring-effect parameters:} Let us clarify the role of the parameter $q^{\star01}$ using Fig. \ref{jump_rate_q} as an example. In Fig. \ref{jump_rate_q_a}, we see a scenario where two ants are positioned at neighboring sites, resulting in the jump rate of the particle in state 1 being $q(\boldsymbol{\eta}) = q^\star$. However, when there are no ants nearby, as shown in Fig. \ref{jump_rate_q_b}, the rate changes to $q(\boldsymbol{\eta}) = q^\star(1 + q^{\star01})$. Therefore, selecting $q^{\star01} < 0$ indicates that the latter rate is lower than the former, suggesting that the presence of ant pairs enhances the rates.

This choice might reflect the behavior observed when ants travel in a line: if an ant behind pushes the leading ant, it can help the leading ant move faster along the trail. This behavior, known as tandem running or tandem carrying, can be seen in some videos of this phenomenon in \cite{videos}. It is a common strategy among certain ant species for the efficient transportation of food or other resources back to the nest. By working together in this way, ants can maximize their collective efficiency and productivity.

The role of $Q^{\star01}$ is similar to that of $q^{\star01}$. Therefore, we refer to both parameters, $q^{\star01}$ and $Q^{\star01}$, as neighboring-effect parameters due to their influence on the rates at which ants locate neighboring sites. We will designate the effect caused by $q^{\star01}$ as the neighboring-effect caused by particle state 1, or as neighboring-pushing type I, and the effect caused by $Q^{\star01}$ as the neighboring-effect caused by particle state 2, or as neighboring-pushing type II.

\subsection{The model stationary distribution}

Let $\boldsymbol{\eta}$ denote a system configuration, defined by position vectors $\textbf{x} = (x_1,...,x_N)$ and state vectors $\textbf{s} = (s_1,...,s_N)$. In addition to the hardcore repulsion, as discussed in \cite{Belitsky2019-1}, we include a short-range interaction energy term in the Hamiltonian, expressed as
\begin{equation}\label{shortrang_energy}
	U(x)=J\sum_{i=1}^{N}\delta_{x_{i}+1,x_{i+1}}
\end{equation}
Here, a positive $J$ indicates repulsion among the particles, whereas it signifies attraction among the ants. Note that the sum $\sum_{i=1}^{N}\delta_{x_{i}+1,x_{i+1}}$ counts the number of pairs of particles located on neighboring sites.

We use $N^\alpha$ to represent the fluctuating number of particles in state $\alpha \in \{1,2\}$, where $N=N^1+N^2$. Additionally, we define the excess as $B(\textbf{s}) = N^1 - N^2$. Therefore, the stationary measure is defined as
\begin{equation}\label{inv_meas}
	\pi(\boldsymbol{\eta})=\dfrac{1}{Z}e^{-\frac{1}{k_{B}T}(U(x)+\lambda B(s))}.
\end{equation}
Here, $Z$ denotes the partition function, defined as $Z=\sum_{\boldsymbol{\eta}}\pi(\boldsymbol{\eta})$. The chemical potential $\lambda$ acts as a Lagrange multiplier to account for fluctuations in the difference between the number of particles in states 1 and 2, referred to as the excess
\begin{equation}
	B(\textbf{s})=\sum_{i=1}^{N}(3-2s_{i}).
\end{equation}
These fluctuations arise from the interaction between the retention of ant pheromones on the trail and their gradual degradation over time. $k_B$ is the Boltzmann constant and $T$ plays a role as temperature.

To facilitate future computational tasks, we introduce
\begin{equation}
	x=e^{\frac{2\lambda}{k_{B}T}}\text{ and }y=e^{\frac{J}{k_{B}T}},
\end{equation}
so that $x>1$ corresponds to an excess of particles in state 1 and repulsive interaction corresponds to $y>1$. Thus, the stationary distribution \eqref{inv_meas} becomes
\begin{equation}\label{inv_meas_1}
	\pi(\boldsymbol{\eta}) = \dfrac{1}{Z}\prod_{i=1}^{N}x^{3/2-s_i}y^{-\delta_{x_{i}+1,x_{i+1}}}.
\end{equation}

Note that the dynamics \eqref{rate_1}–\eqref{rate_3} and stationary measure \eqref{inv_meas_1} of the model are described in parameter form. Parameters cannot be arbitrarily chosen to ensure that the process, governed by the dynamics, possesses a measure in the form of \eqref{inv_meas} as its invariant distribution. Namely, the model’s parameters must conform to the following constraints:

\begin{align}
	x&=\dfrac{q^{\star}}{f^{\text{\ensuremath{\star}}}},\label{cond1}\\
	y&=\dfrac{1+\dfrac{1}{f^{\star}}Q^{\star}}{1+q^{\star01}+\dfrac{1}{f^{\star}}Q^{\star}(1+Q^{\star01})},\label{cond_y}\\
	f^{1\star}&=\dfrac{x}{1+x}-\dfrac{x}{1+x}\dfrac{Q^{\star}}{q^{\star}}-1,\\
	f^{\star1}&=\dfrac{1}{1+x}+\dfrac{x}{1+x}\dfrac{Q^{\star}}{q^{\star}}-1,\\
	f^{10\star}&=\dfrac{1}{1+x}(1+q^{\star01})+\dfrac{x}{1+x}\dfrac{Q^{\star}}{q^{\star}}(1+Q^{\star01})-\dfrac{1}{1+x}-\dfrac{x}{1+x}\dfrac{Q^{\star}}{q^{\star}},\\
	f^{\star01}&=\dfrac{x}{1+x}(1+q^{\star01})-\dfrac{x}{1+x}\dfrac{Q^{\star}}{q^{\star}}(1+Q^{\star01})-\dfrac{x}{1+x}+\dfrac{x}{1+x}\dfrac{Q^{\star}}{q^{\star}}.\label{cond2}
\end{align}
In other words, one can summarize the results by Theorem \ref{main_theorem} below. The proof of this theorem can be found in Appendix \ref{condition_of_model}.

\begin{theorem}\label{main_theorem}
	If the parameters specified in rates \eqref{rate_1}--\eqref{rate_3} satisfy the conditions outlined in \eqref{cond1}--\eqref{cond2}, the invariant measure of the process is given by
	\begin{equation}\label{inv_meas_theorem}
		\hat{\pi}(\boldsymbol{\eta}) = \dfrac{1}{Z}\left(\dfrac{q^{\star}}{f^{\text{\ensuremath{\star}}}}\right)^{\sum_{i=1}^{N}-3/2+ s_i}\left( \dfrac{1+\dfrac{1}{f^{\star}}Q^{\star}}{1+q^{\star01}+\dfrac{1}{f^{\star}}Q^{\star}(1+Q^{\star01})} \right)^{-\sum_{i=1}^{N}\delta_{x_{i+1}, x_i+1}}
	\end{equation}
	where $Z$ denotes the partition function.
	
\end{theorem}

\begin{remark}
	The parameters $q^\star$, $Q^\star$, $f^\star$, $q^{\star01}$, and $Q^{\star01}$ play a crucial role in the dynamics of the model, as the other parameters can be expressed through them.
\end{remark}
\begin{remark}
Similar to Chowdhury et al.'s model with random-sequential updating, simulations show that the loose cluster property does not hold in our model.
\end{remark}

\section{Average current and velocity}\label{current_velocity}
In this section, we calculate the average current and velocity of ants. Before doing so, we compute certain quantities such as the average stationary densities and the average dwell times of particles in the two states. These calculations help compute  the average current and velocity.

\subsection{Average dwell times and stationary densities}
\textbf{Average excess:} As highlighted in \cite{Belitsky2019-1}, the most straightforward measure for defining particle distribution is the average excess density of type 1 over type 2, represented as
\begin{equation}
	\sigma = \dfrac{\left< N^1 \right> - \left< N^2 \right>}{L}.
\end{equation}
Because we use the grand-canonical ensemble  \eqref{grand_can_ensem}, which is of the same form as the one in \cite{Belitsky2019-1}, we similarly obtain the value $\sigma$ as follows

\begin{equation}\label{averexcess}
	\sigma = \dfrac{1-x}{1+x}\rho_p.
\end{equation}
where $\rho_p$ is the particle density, i.e., $\rho_p = N/L$. However, the difference in the value of $\sigma$ compared to the one stated in \cite{Belitsky2019-1} arises from the value $x$, as our model's dynamics are not the same as those in the paper. \\

\noindent\textbf{Average densities:} We denote by
\begin{equation}
	\rho^\alpha := \left< \delta_{s_i,\alpha} \right> = \dfrac{1}{L}\left< N^\alpha \right>,\ \ \alpha \in \{1,2\},
\end{equation}
the average densities of particles in states $1$ and $2$. Since $\rho^1 +\rho^2 = \rho_p$ and using \eqref{averexcess}, we obtain
\begin{equation}\label{statedensity}
	\rho^1 =\dfrac{1}{1+x}\rho_p, \ \ \rho^2 = \dfrac{x}{1+x}\rho_p.
\end{equation}
\vspace{0.2cm}

\noindent\textbf{Average dwell times:} Let $\tau^{\alpha}$, where $\alpha \in \{1,2\}$, denote the average dwell time that a particle spends in states 1 and 2, respectively. Due to ergodicity, we have $\tau^\alpha = \rho^\alpha/\rho_p$. This yields the following formulas for average dwell times
\begin{equation}\label{dwelltime}
	\tau^1 = \dfrac{1}{1+x}, \ \ \tau^2 = \dfrac{x}{1+x}.
\end{equation}
From equations \eqref{statedensity} and \eqref{dwelltime}, we derive the balance equation
\begin{equation}
	\dfrac{\rho^1}{\rho^2} = \dfrac{\tau^1}{\tau^2} =\dfrac{f^\star}{q^\star}.
\end{equation}
This equation expresses the ensemble ratio in relation to the single-particle translocation rates $q^\star$ and the single-particle transition rate $f^\star$.

\subsection{Average stationary current and velocity}
As previously noted, we analyze the collective movement of ants by studying the motion of particles governed by dynamics \eqref{rate_1}--\eqref{rate_3}. The current $j_p$, quantified as the average number of particles traversing a site per unit time, reflects the collective motion of these particles. Additionally, the average speed $\nu_p$ of particles is another characteristic of collective movement, which is linked to $j_p$ through the equation
\begin{equation}
	j_p =\rho_p \nu_p.
\end{equation}

Instead of using the form of rates \eqref{rate_1}--\eqref{rate_2} to compute $j_p$ and $\nu_p$, it is equivalent to use the rates in headway form \eqref{rate1_h}--\eqref{rate2_h} for convenience. The advantage is that it incorporates the headway distribution \eqref{headway_distribution} into the computation. Namely, one has
\begin{equation}\label{particle_current}
	j_p =  \langle \delta_{s_i,1}(q^\star (1+q^{\star01})\theta_{i}^1+q^\star \theta_{i}^{>1}) \rangle+\langle \delta_{s_i,2}(Q^\star (1+Q^{\star01})\theta_{i}^1+Q^\star \theta_{i}^{>1}) \rangle. 
\end{equation}
Here, $\langle\cdot \rangle$ indicates taking the expectation with respect to the distribution \eqref{grand_can_ensem}, and the headway variable $\theta_i^{>1}$ indicates that there is more than one empty site ahead of particle $i$, i.e., $x_{i+1}-x_i >2$. By utilizing the factorization property of the grandcanonical ensemble \eqref{grand_can_ensem} and the identities in \eqref{statedensity}, one obtains
\begin{align}
	j_p & =  \langle \delta_{s_i,1} \rangle \langle(q^\star (1+q^{\star01})\theta_{i}^1+q^\star \theta_{i}^{>1}) \rangle+\langle \delta_{s_i,2}\rangle \langle(Q^\star (1+Q^{\star01})\theta_{i}^1+Q^\star \theta_{i}^{>1}) \rangle\nonumber\\
	&= \dfrac{1}{1+x}\rho_p \big(q^\star (1+  q^{\star01})\langle \theta_{i}^1 \rangle +q^\star \langle \theta_{i}^{>1} \rangle\big)  +\dfrac{x}{1+x}\rho_p \big(Q^\star (1+Q^{\star01})\langle\theta_{i}^1\rangle+Q^\star \langle\theta_{i}^{>1}\rangle\big).
\end{align}
From headway distribution \eqref{headway_distribution}, one can compute $\langle \theta_i^0 \rangle, \langle \theta_i^1 \rangle$ and one has
\begin{equation}
	\langle \theta_i^{>1} \rangle  = 1- \langle \theta_i^{0} \rangle - \langle \theta_i^{1} \rangle.
\end{equation}

\begin{figure}[H]
	\centering
	\begin{subfigure}[b]{0.4\textwidth}
			\centering
			\includegraphics[width=\textwidth]{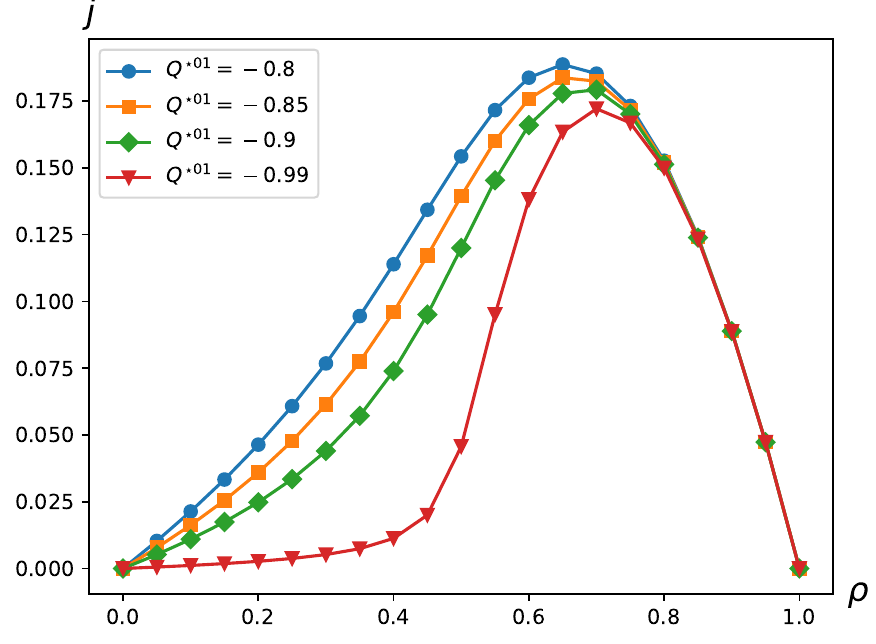}
			\caption{Average current $j$}
			\label{fig:j1}
		\end{subfigure}
	\begin{subfigure}[b]{0.4\textwidth}
			\centering
			\includegraphics[width=\textwidth]{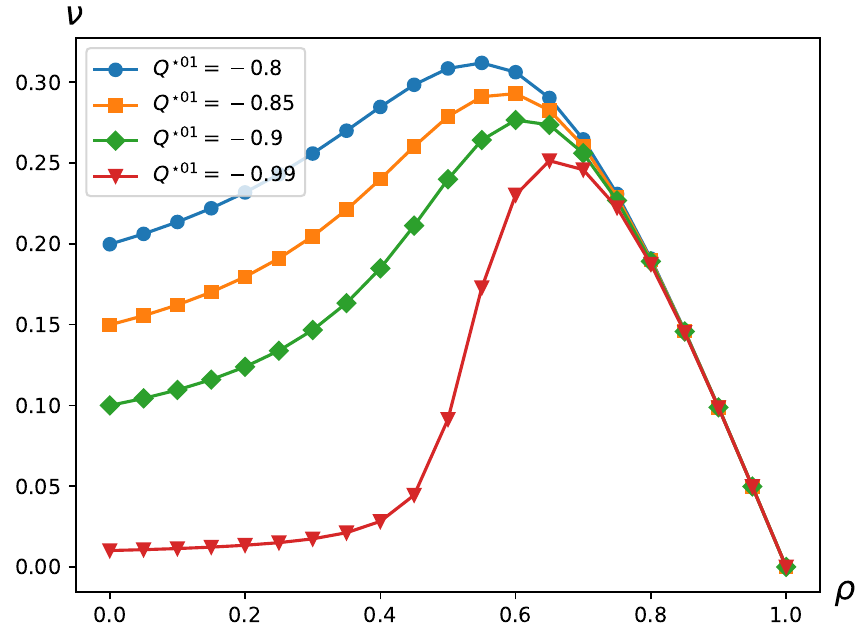}
			\caption{Average velocity $\nu$}
			\label{fig:v1}
		\end{subfigure}
	\caption{Average current $j$ and velocity $\nu$ as functions of ant density $\rho$. $Q^\star = 1, q^\star = 0.25, q^{\star01}=-0.7, f^\star = 0.0005$ with different values of $Q^{\star01} = -0.8; -0.85; -0.9; -0.99$.}
	\label{fig:jv1}
\end{figure}

\begin{figure}[H]
	\centering
	\begin{subfigure}[b]{0.4\textwidth}
		\centering
		\includegraphics[width=\textwidth]{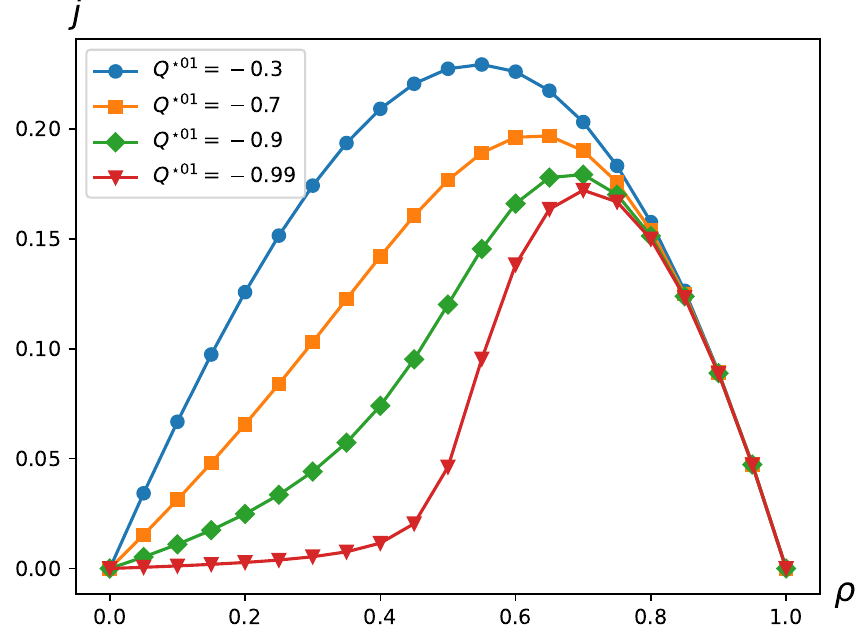}
		\caption{Average current $j$}
		\label{fig:j2}
	\end{subfigure}
	\begin{subfigure}[b]{0.4\textwidth}
		\centering
		\includegraphics[width=\textwidth]{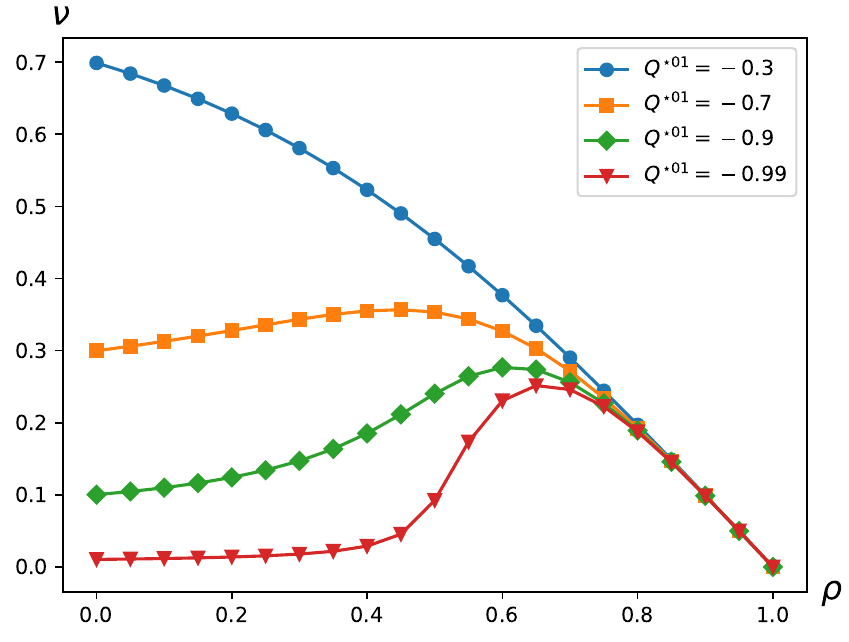}
		\caption{Average velocity $\nu$}
		\label{fig:v2}
	\end{subfigure}
	\caption{Average current $j$ and velocity $\nu$ as functions of ant density $\rho$. $Q^\star = 1, q^\star = 0.25, q^{\star01}=-0.2, f^\star = 0.0005$ with different values of $Q^{\star01} = -0.3; -0.7; -0.9; -0.99$.}
	\label{fig:jv2}
\end{figure}

\begin{figure}[H]
	\centering
	\begin{subfigure}[b]{0.4\textwidth}
		\centering
		\includegraphics[width=\textwidth]{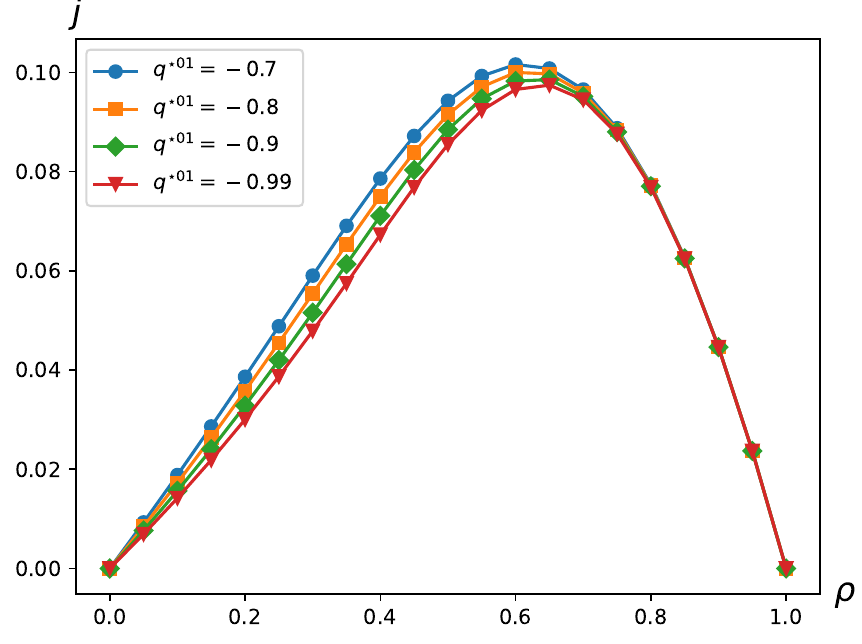}
		\caption{Average current $j$}
		\label{fig:j3}
	\end{subfigure}
	\begin{subfigure}[b]{0.4\textwidth}
		\centering
		\includegraphics[width=\textwidth]{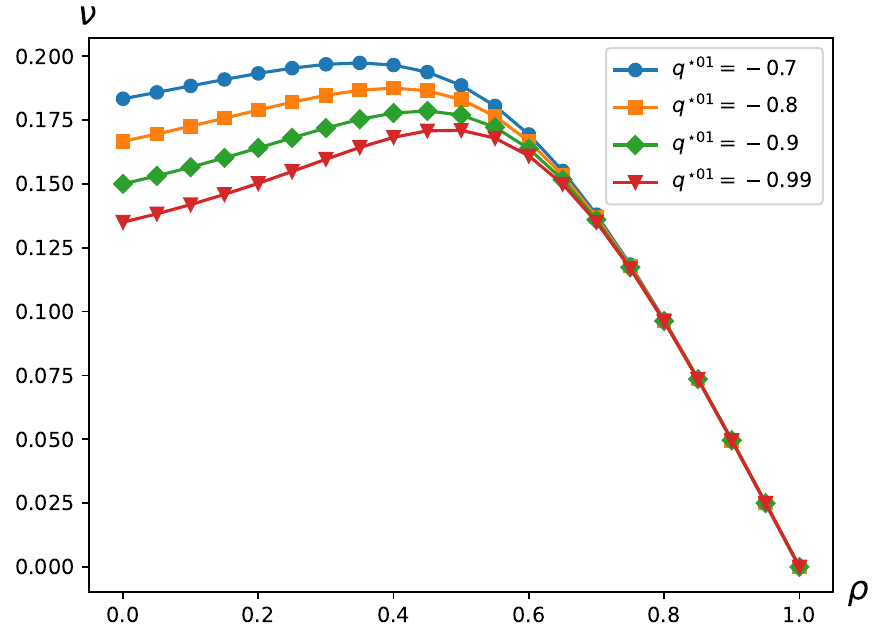}
		\caption{Average velocity $\nu$}
		\label{fig:v3}
	\end{subfigure}
	\caption{Average current $j$ and velocity $\nu$ as functions of ant density $\rho$. $Q^\star = 1, q^\star = 0.25, Q^{\star01}=-0.6, f^\star = 0.5$ with different values of $q^{\star01} = -0.7; -0.8; -0.9; -0.99$.}
	\label{fig:jv3}
\end{figure}

\begin{figure}[H]
	\centering
	\begin{subfigure}[b]{0.4\textwidth}
		\centering
		\includegraphics[width=\textwidth]{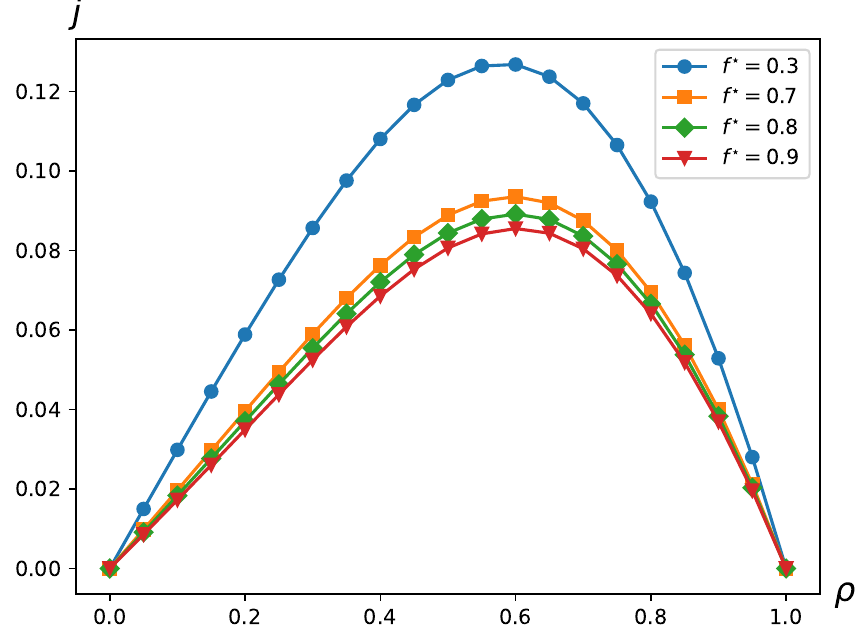}
		\caption{Average current $j$}
		\label{fig:j4}
	\end{subfigure}
	\begin{subfigure}[b]{0.4\textwidth}
		\centering
		\includegraphics[width=\textwidth]{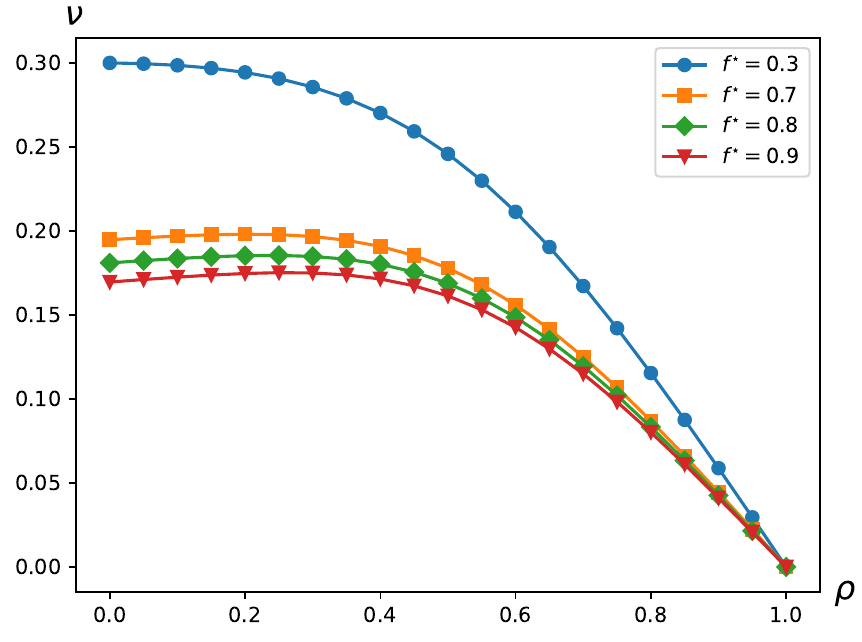}
		\caption{Average velocity $\nu$}
		\label{fig:v4}
	\end{subfigure}
	\caption{Average current $j$ and velocity $\nu$ as functions of ant density $\rho$. $Q^\star = 1, q^\star = 0.25, Q^{\star01}=-0.4, q^{\star01} = -0.8$ with different values of $f^{\star} = 0.3; 0.7; 0.8; 0.9$.}
	\label{fig:jv4}
\end{figure}

\begin{figure}[H]
	\centering
	\begin{subfigure}[b]{0.4\textwidth}
		\centering
		\includegraphics[width=\textwidth]{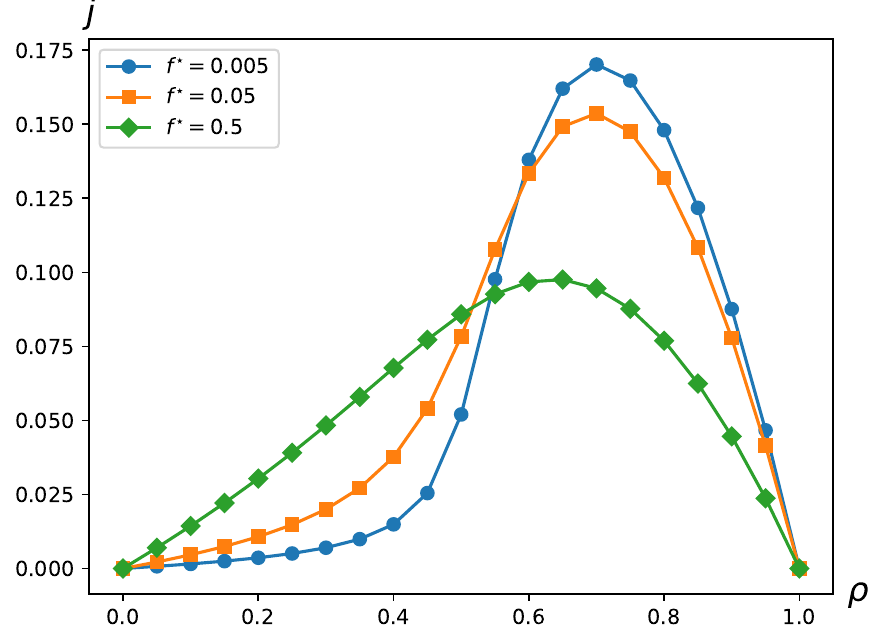}
		\caption{Average current $j$}
		\label{fig:j11}
	\end{subfigure}
	\begin{subfigure}[b]{0.4\textwidth}
		\centering
		\includegraphics[width=\textwidth]{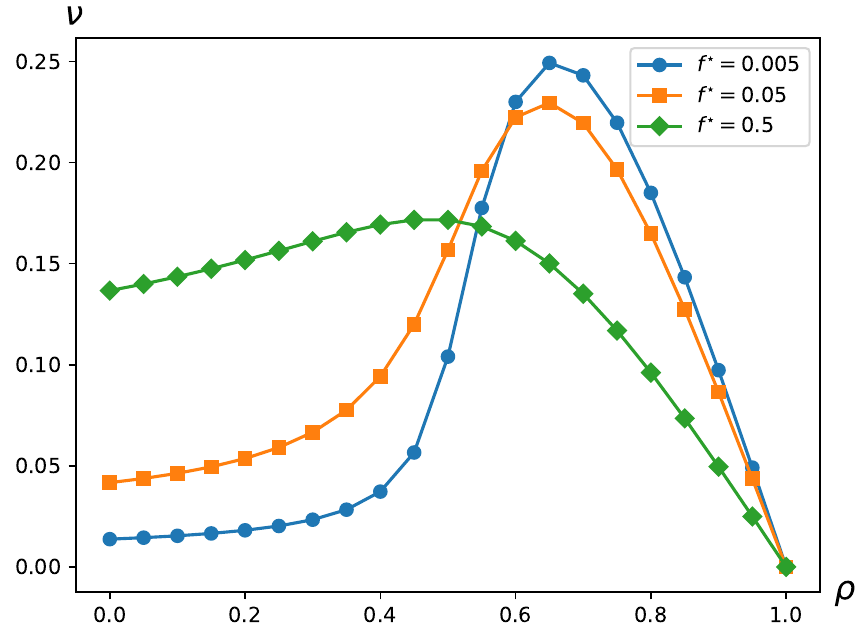}
		\caption{Average velocity $\nu$}
		\label{fig:v11}
	\end{subfigure}
	\caption{Average current $j$ and velocity $\nu$ as functions of ant density $\rho$. $Q^\star = 1, q^\star = 0.25, q^{\star01}=-0.2, Q^{\star01}=-0.99$ with different values of $f^{\star} = 0.005; 0.05; 0.5$.}
	\label{fig:jv11}
\end{figure}

From this point, we have sufficient ingredients to draw graphs showing  the average current and velocity of ants. However, notice that the average current $j$ and velocity $v$ of ants with density $\rho = 1 - \rho_p$ are related to $j_p$ and $\nu_p$ as follows
\begin{equation}\label{current_velocity_of_ants}
	\begin{cases}
		j(\rho) = j_p(1-\rho_p),\\
		\nu(\rho) = j(\rho)/\rho.
	\end{cases}
\end{equation}
The influence of neighboring-effect parameters $q^{\star01}, Q^{\star01}$, and evaporating rate $f^\star$ on the average current and velocity of ants is examined. Two scenarios are considered: when $Q^{\star01} \leq q^{\star01}$ and when $Q^{\star01} > q^{\star01}$. Refer to Figs. \ref{fig:jv1} and \ref{fig:jv2}  for the former case, and Figs. \ref{fig:jv3} and \ref{fig:jv4} for the latter.  See Fig. \ref{fig:jv11} for the influence of the evaporating rate $f^\star$ on the average current and velocity of ants. In the following graphs, we select the values of $Q^\star$ and $q^\star$ such that $Q^\star > q^\star$, indicating that the rate with pheromone at the destination site exceeds the rate without pheromone at the site. Namely, in all the graphs presented below, the values of $Q^\star$ and $q^\star$ are set to 1 and 0.25, respectively.

\section{Discussion}\label{discussion}
For clarity in the following discussion, it is important to note that the non-monotonic trend in the average velocity of ants $\nu(\rho)$ \eqref{current_velocity_of_ants} occurs when it reaches an intermediate maximum at a density $\rho_m$ within the range of $(0,1)$. This trend manifests only when the interaction strength $y$ \eqref{cond_y} exceeds a critical value $y_c := 2$. Specifically, when $y > 2$, the stationary average velocity of ants achieves its maximum value of $\dfrac{(f^\star + Q^\star) q^\star y}{4(f^\star + q^\star)(-1 + y)}$ at a density of $\rho_m = \dfrac{-4+2y}{-4+3y} \in (0,1)$.
	
To better understand how dynamics at the microscopic scale influence this non-monotonic trend, as previously discussed, we examine the effects of neighboring-effect parameters $Q^{\star01}$, $q^{\star01}$, and the evaporation rate $f^\star$ on the average current and velocity of ants in two scenarios: when $Q^{\star01} \leq q^{\star01}$ and when $Q^{\star01} > q^{\star01}$. Refer to Figs. \ref{fig:jv1} to \ref{fig:jv2} for the former case and Figs. \ref{fig:jv3} and \ref{fig:jv4} for the latter. 
	
Recall that we assume $q^{\star01}$ and $Q^{\star01}$ lie within the range $(-1, 0)$, indicating that a trailing ant pushes the leading ant located at its neighboring site. This pushing effect caused by $q^{\star01}$ and $Q^{\star01}$ is categorized as neighboring-pushing type I and type II, respectively. We classify these types of neighboring-pushing as strong or weak depending on the values of $q^{\star01}$ and $Q^{\star01}$ and their proximity to $-1$ or $0$, respectively.

In the scenario where $Q^{\star01} \leq q^{\star01}$, the condition $y > 2$ holds under two specific scenarios. Firstly, both neighboring-pushing effects are strong, specifically when $-1 < Q^{\star01} < -\frac{1}{2}$ and $Q^{\star01} \leq q^{\star01} \leq -1 - Q^{\star01}$. Secondly, if neighboring-pushing type I is not sufficiently strong, meaning $-1 - Q^{\star01} < q^{\star01}$, then neighboring-pushing type II must be strong ($-1 < Q^{\star01} < -\frac{1}{2}$), and the evaporation rate $f^\star$ must be sufficiently small ($f^\star < -\dfrac{1 + 2Q^{\star01}}{1 + 2q^{\star01}}$). Refer to Fig. \ref{fig:jv1} for an illustration of the first scenario. Fig. \ref{fig:jv2} depicts the second scenario, except when $Q^{\star01} = -0.3$, where the expected non-monotonic trend does not appear due to the failure of the last condition, i.e., $f^\star < -\dfrac{1 + 2Q^{\star01}}{1 + 2q^{\star01}}$ not being satisfied.

Thus, in this case ($Q^{\star01} \leq q^{\star01}$), the non-monotonic trend occurs under two specific conditions: either both types of neighboring-pushing are sufficiently strong, or if neighboring-pushing type II is strong while neighboring-pushing type I is weaker, the rate of pheromone evaporation $f^\star$ must be sufficiently small to enhance the overall speed of the ants. In the latter scenario, the slow evaporation of pheromones compensates for the weaker neighboring-pushing type I, contributing to the overall speed enhancement of the ants.

In the case where $Q^{\star01} > q^{\star01}$, the condition $y > 2$ holds only if one of two scenarios is true. First, if both neighboring-effect parameters are strong, specifically $-1 < q^{\star01} < Q^{\star01} < -\dfrac{1}{2}$. Refer to Fig. \ref{fig:jv3} for an illustration of this scenario. Second, if the neighboring-effect parameter $Q^{\star01}$ is not strong enough, namely $-\dfrac{1}{2} < Q^{\star01} < 0$, then $q^{\star01}$ must be strong ($-1 < q^{\star01} < -1 - Q^{\star01}$) and the rate of evaporation $f^\star$ must be large enough ($f^\star > -\dfrac{1 + 2Q^{\star01}}{1 + 2q^{\star01}}$).

However, the latter scenario appears inconsistent with reality, as pheromones typically dissipate slowly \cite{Schadschneider2011}. Nevertheless, our goal is to explain why the non-monotonic trend occurs in this context. The strong influence of neighboring-pushing type I significantly enhances the speed of ants. Moreover, when the rate of pheromone evaporation $f^\star$ is high, there is a rapid transition of many particles from state 2 to state 1. As a result, neighboring-pushing type I, driven by particles in state 1, becomes more frequently activated, which contributes to the occurrence of the non-monotonic trend in the average velocity of ants.

Fig. \ref{fig:jv4} depicts the second scenario, except when $f^{\star} = 0.3$, where the expected non-monotonic trend does not appear because the last condition is not met; specifically, the condition $f^\star > -\dfrac{1 + 2Q^{\star01}}{1 + 2q^{\star01}}$ is not satisfied. Additionally, as shown in Fig. \ref{fig:jv4}, the non-monotonic trend is not clearly observable in this situation.

Another interesting phenomenon is the impact of the pheromone evaporation rate, $f^\star$, on the average current and velocity of ants. Fig. \ref{fig:jv11} illustrates that at low ant density, both the current and velocity are higher with a higher $f^\star$ value. Conversely, at high density, both the current and velocity decrease as $f^\star$ increases.

In the limiting case where $Q^{\star01}$ and $q^{\star01}$ approximate zero, i.e., there is no neighboring-effect. In this case, one has $y \approx 1$, and we consider this case is non-interacting, our model simplifies to TASEP \eqref{TSEP} with different hopping probabilities. Specifically, in this case, the average current and velocity of ants in \eqref{current_velocity_of_ants} become
\begin{align}\label{current_y_1}
	\nu = \dfrac{q^\star + Q^\star x}{1+x}(1- \rho),\  \ j = \dfrac{q^\star + Q^\star x}{1+x}\rho(1 - \rho).
\end{align}
Note that the quantities $\nu$ and $j$ in \eqref{current_y_1} differ from those in \eqref{current_d0} of the TASEP only by the factor
 $\dfrac{q^\star + Q^\star x}{1+x}$.
 
Let us consider another limiting case where $f^\star \approx 0$. In this scenario, the dynamics of our model \eqref{rate_1}--\eqref{rate_3} apply only to $Q_i(\boldsymbol{\eta})$. This is because, when $f^\star \approx 0$, the transition rate $f_i(\boldsymbol{\eta})$ is nearly zero, and the density of particles in state 2, $\rho^2$ \eqref{statedensity}, is 1. This implies that the ant trail is saturated with pheromones. Consequently, the jump rate of type 1 particles, $q_i(\boldsymbol{\eta})$, is rarely active, since almost all particles are in state 2. Under these conditions, our model aligns with the model considered in \cite{Antal2000}, specifically, TASEP with next-nearest-neighbor interaction. In particular, the corresponding rates $r$ and $q$ in equations (2) and (3) of that paper correspond to $Q^\star$ and $Q^\star(1 + Q^{\star 01})$, respectively. In this case, the average current and velocity of ants \eqref{current_velocity_of_ants} are given by:
\begin{equation}
	\begin{cases}
		j(\rho) &= Q^\star (1-\rho)\left(1+\dfrac{1-\sqrt{1+4\rho(1-\rho)Q^{\star01}}}{2\rho Q^{\star01}}\right),\\
		\nu(\rho) &= j(\rho)/\rho.
	\end{cases}
\end{equation}
Moreover, in this case, we have $y = \dfrac{1}{1+Q^{\star 01}}$. Thus, a non-monotonic trend in the average velocity of ants appears only if $-1 < Q^{\star 01} < -\dfrac{1}{2}$.

\section{Conclusions}\label{conclusion}
We have presented an exactly solvable model for single-lane unidirectional ant traffic, based on the dynamics proposed by Chowdhury et al. However, our model utilizes random-sequential updating instead of the parallel updating used in the original model. Additionally, we introduced neighboring-effect parameters to describe the tandem running behavior of ants.

Our exactly solvable model allows us to identify conditions under which the non-monotonic trend of collective ant velocity appears. Interestingly, our model reveals a phenomenon not mentioned in Chowdhury et al.'s work: at low densities, the average current and velocity can exceed those at higher probabilities of evaporation. This behavior reverses at high densities, where smaller probabilities of evaporation yield greater average current and velocity.

However, unlike Chowdhury et al.'s model using parallel updating, our model does not exhibit the loose cluster property due to the use of random-sequential updating. This finding aligns with Chowdhury et al.'s results when applying the same updating rule, i.e., random-sequential updating.

\section*{Appendices}
\appendix
\renewcommand{\thesection}{\Alph{section}}

\section{Stationary conditions}\label{condition_of_model}

We will outline a method for determining the conditions on the parameters present in the rates \eqref{rate_1}--\eqref{rate_3} to ensure that the process's invariant measure aligns with the form described in \eqref{inv_meas}, i.e., we will show how to obtain the constraints \eqref{cond1}--\eqref{cond2}. To accomplish this, we will utilize the approach introduced in \cite{Belitsky2019-1}. First, we will map the process onto a corresponding headway structure. Then, by using a discrete version of Noether's theorem \eqref{local_divergence}, we can rewrite the master equation of the process into a local divergence form when the system reaches equilibrium. From this point, one can determine the constraints.
\subsection{Master equation}
We represent the probability of the system being in configuration being in configuration $\boldsymbol{\eta}$ at time $t$ as $\mathbb{P}(\boldsymbol{\eta},t)$. 
The evolution of $\mathbb{P}(\boldsymbol{\eta},t)$ over time is governed by the master equation
\begin{equation}\label{master_eq}
	\begin{split}
		\dfrac{d}{dt}\mathbb{P}(\boldsymbol{\eta},t)=\sum_{i=1}^{N}\bigg(q_{i}(\boldsymbol{\eta}_{1}^{i})\mathbb{P}(\boldsymbol{\eta}_{1}^{i},t)+&Q_{i}(\boldsymbol{\eta}_{2}^{i})\mathbb{P}(\boldsymbol{\eta}_{2}^{i},t)+f_{i}(\boldsymbol{\eta}_{3}^{i})\mathbb{P}(\boldsymbol{\eta}_{3}^{i},t)\\&-(q_{i}(\boldsymbol{\eta})+Q_{i}(\boldsymbol{\eta})+f_{i}(\boldsymbol{\eta}))\mathbb{P}(\boldsymbol{\eta},t)\bigg).
	\end{split}
\end{equation}
Here, $\boldsymbol{\eta}_{1}^{i}$ represents the configuration leading to $\boldsymbol{\eta}$ before the forward translocation of particle $i$ in state 1 (with $x_{i}^{1}=x_{i}-1$ and $s_{i}^{1}=3-s_{i}$), $\boldsymbol{\eta}_{2}^{i}$ signifies the configuration leading to $\boldsymbol{\eta}$ before the forward translocation of particle $i$ in state 2 ($x_{i}^{2}=x_{i}-1$ and $s_{i}^{2}=s_{i}$), and $\boldsymbol{\eta}_{3}^{i}$ denotes the configuration leading to $\boldsymbol{\eta}$ before the pheromone at site $x_i$ evaporates ($x_{i}^{3}=x_{i}$ and $s_{i}^{3}=3-s_{i}$).

At equilibrium, the master equation \eqref{master_eq} does not depend on time, i.e., $\dfrac{d}{dt}\mathbb{\pi}(\boldsymbol{\eta},t) = 0$. Thus,\begin{equation}\label{master_eq1}
	\sum_{i=1}^{N}\big(q_{i}(\boldsymbol{\eta}_{1}^{i})\pi(\boldsymbol{\eta}_{1}^{i})+Q_{i}(\boldsymbol{\eta}_{2}^{i})\pi(\boldsymbol{\eta}_{2}^{i})+f_{i}(\boldsymbol{\eta}_{3}^{i})\pi(\boldsymbol{\eta}_{3}^{i})-(q_{i}(\boldsymbol{\eta})+Q_{i}(\boldsymbol{\eta})+f_{i}(\boldsymbol{\eta}))\pi(\boldsymbol{\eta})\big) = 0.
\end{equation}
Dividing both sides of \eqref{master_eq1}  by  $\pi(\boldsymbol{\eta})$, one obtains
\begin{equation}
	\sum_{i=1}^{N}\big(q_{i}(\boldsymbol{\eta}_{1}^{i})\dfrac{\pi(\boldsymbol{\eta}_{1}^{i})}{\pi(\boldsymbol{\eta})}+Q_{i}(\boldsymbol{\eta}_{2}^{i})\dfrac{\pi(\boldsymbol{\eta}_{2}^{i})}{\pi(\boldsymbol{\eta})}+f_{i}(\boldsymbol{\eta}_{3}^{i})\dfrac{\pi(\boldsymbol{\eta}_{3}^{i})}{\pi(\boldsymbol{\eta})}-(q_{i}(\boldsymbol{\eta})+Q_{i}(\boldsymbol{\eta})+f_{i}(\boldsymbol{\eta}))\big)=0.\label{mas_eq2}
\end{equation}
For convenience in writing, let us denote the following quantities
\begin{align}
	D_{i}(\boldsymbol{\eta)} & =q_{i}(\boldsymbol{\eta}_{1}^{i})\dfrac{\pi(\boldsymbol{\eta}_{1}^{i})}{\pi(\boldsymbol{\eta})}-q_{i}(\boldsymbol{\eta}),\\
	E_{i}(\boldsymbol{\eta)} & =Q_{i}(\boldsymbol{\eta}_{2}^{i})\dfrac{\pi(\boldsymbol{\eta}_{2}^{i})}{\pi(\boldsymbol{\eta})}-Q_{i}(\boldsymbol{\eta}),\\
	F_{i}(\boldsymbol{\eta}) & =f_{i}(\boldsymbol{\eta}_{3}^{i})\dfrac{\pi(\boldsymbol{\eta}_{3}^{i})}{\pi(\boldsymbol{\eta})}-f_{i}(\boldsymbol{\eta}).
\end{align}
Because of the periodicity, the master equation $\eqref{mas_eq2}$ is satisfied if one has 
\begin{equation}\label{local_divergence}
	D_{i}(\boldsymbol{\eta})+E_{i}(\boldsymbol{\eta})+F_{i}(\boldsymbol{\eta})=\Phi_{i}(\boldsymbol{\eta})-\Phi_{i+1}(\boldsymbol{\eta}).
\end{equation}
This condition applies to all configurations of the system, characterized by a set of functions $\Phi_{i}(\boldsymbol{\eta})$ that adhere to the condition $\Phi_{N+1}(\boldsymbol{\eta)}=\Phi_{1}(\boldsymbol{\eta})$. One can consider the lattice divergence condition \eqref{local_divergence} as Noether's theorem in discrete form. The specific forms of the functions $\Phi_i$ will be detailed in the subsequent subsection.

\subsection{Mapping to the headway process}
In addition to specifying a system's configuration $\boldsymbol{\eta} = (\eta_1,...,\eta_L)$ by its position and state vectors $\textbf{x} = (x_1,...,x_N)$ and $\textbf{s} = (s_1,...,s_N)$ respectively, it can also be determined by the state vector along with the headway vector $\textbf{m} = (m_1,...,m_N)$. Here, $m_i$ represents the number of empty sites between the $i$-th and $(i+1)$-th particles, i.e., one has $m_i = x_{i+1} - (x_i + 1) \mod L$.

To represent a headway of length $r$, instead of using $\delta_{m_i,r}$, we adopt the concise notation $\theta_{i}^r$. Specifically, we define
\begin{equation}\label{newvar}
	\theta_{i}^r := \delta_{m_i, r} = \delta_{x_{i+1}, x_i + 1 +r}.
\end{equation}
Again, concerning the headway variable, the index $i$ is also considered modulo $N$, implying $\theta_0^r \equiv \theta_N^r$.

Because of steric hard-core repulsion, the $i$-th particle undergoes a forward translocation from $x_{i}$ to $x_{i}+1$, corresponding to the transition $(m_{i-1},m_{i})\to(m_{i-1}+1,m_{i}-1)$, whenever $m_{i}>0$. Expressed in the new stochastic variables $\boldsymbol{\zeta}=(\textbf{m},\textbf{s})$, where $\textbf{m}$ represents the distance vector and $\textbf{s}$ is the state vector, the invariant distribution \eqref{inv_meas_1} can be reformulated as
\begin{equation}\label{inv_meas_2}
	\tilde{\pi}(\boldsymbol{\zeta}) = \dfrac{1}{\tilde{Z}} \prod_{i=1}^{N}x^{3/2-s_i}y^{-\theta_{i}^0},
\end{equation} 
where $\tilde{Z}$ is the partition function. Additionally, the transition rates  \eqref{rate_1}--\eqref{rate_3} can be rewritten as
\begin{align}
	\tilde{q}_{i}(\boldsymbol{\zeta}) & =q^{\star}\delta_{s_{i},1}(1+q^{\star01}\theta_{i}^{1})(1-\theta_{i}^{0}),\label{rate1_h}\\
	\tilde{Q}_{i}(\boldsymbol{\zeta}) & =Q^{\star}\delta_{s_{i},2}(1+Q^{\star01}\theta_{i}^{1})(1-\theta_{i-1}^{0}),\label{rate2_h}\\
	\tilde{f}_{i}(\boldsymbol{\zeta}) & =f^{\star}\delta_{s_{i},2}(1+f^{1\star}\theta_{i-1}^{0}+f^{\star1}\theta_{i}^{0}+f^{10\star}\theta_{i-1}^{1}+f^{\star01}\theta_{i}^{1}).\label{rate3_h}
\end{align}
Given configuration $\boldsymbol{\zeta}$, let $\boldsymbol{\zeta}^{i,1}$ and $\boldsymbol{\zeta}^{i,2}$ represent configurations resulting from forward translocation when the state of particle $i$ is 1 and 2, respectively. Additionally, let $\boldsymbol{\zeta}^{i,3}$ denote the configuration leading to $\boldsymbol{\zeta}$ before pheromone evaporation at site $x_i$. Thus, configurations $\boldsymbol{\zeta}^{i,1}$, $\boldsymbol{\zeta}^{i,2}$, and $\boldsymbol{\zeta}^{i,3}$ are defined through the distance and state vectors by
\begin{align}
	m^{i,1}_j := m_j + \delta_{j,i-1} - \delta_{j,i}\ &\text{and }\  s^{i,1}_j := s_j + (3-2s_j) \delta_{j,i},\\
	m^{i,2}_j := m_j + \delta_{j,i-1} - \delta_{j,i}\ &\text{and }\  s^{i,2}_j := s_j,\\
	m_j^{i,3} := m_j\ &\text{and }\  s_j^{i,3}  := s_j + (3-2s_j)\delta_{j,i}.
\end{align}
This yields the master equation for the headway process
\begin{equation}\label{mas_eq3}
	\dfrac{d\mathbb{P}(\boldsymbol{\zeta},t)}{dt} = \sum_{i=1}^{N}H_i(\boldsymbol{\zeta},t)
\end{equation}
with
\begin{align}
	\begin{split}
		H_i(\boldsymbol{\zeta},t) =  \tilde{q}_i(\boldsymbol{\zeta}^{i,1})\mathbb{P}(\boldsymbol{\zeta}^{i,1},t) &- \tilde{q}_i(\boldsymbol{\zeta})\mathbb{P}(\boldsymbol{\zeta},t) 
		+ \tilde{Q}_i(\boldsymbol{\zeta}^{i,2})\mathbb{P}(\boldsymbol{\zeta}^{i,2},t) \\- \tilde{Q}_i(\boldsymbol{\zeta})\mathbb{P}(\boldsymbol{\zeta},t) 
		& + \tilde{f}_i(\boldsymbol{\zeta}^{i,3})\mathbb{P}(\boldsymbol{\zeta}^{i,3},t) - \tilde{f}_i(\boldsymbol{\zeta})\mathbb{P}(\boldsymbol{\zeta},t) 
	\end{split}
\end{align}
where
\begin{align}
	\tilde{q}_{i}(\boldsymbol{\zeta}^{i,1}) & =q^{\star}\delta_{s_{i},2}(1+d^{\star01}\theta_{i}^{0})(1-\theta_{i-1}^{0}),\\
	\tilde{Q}_{i}(\boldsymbol{\zeta}^{i,2}) & =Q^{\star}\delta_{s_{i},2}(1+Q^{\star01}\theta_{i}^{0})(1-\theta_{i-1}^{0}),\\
	\tilde{f}_{i}(\boldsymbol{\zeta}^{i,3}) & =f^{\star}\delta_{s_{i},1}(1+f^{1\star}\theta_{i-1}^{0}+f^{\star1}\theta_{i}^{0}+f^{10\star}\theta_{i-1}^{1}+f^{\star01}\theta_{i}^{1}).
\end{align}
Thus, at equilibrium, the master equation \eqref{mas_eq3} yields
\begin{equation}\label{mas_eq4}
	\begin{split}
		\sum_{i=1}^{N}\bigg(\tilde{q}_i(\boldsymbol{\zeta}^{i,1})\tilde{\pi}(\boldsymbol{\zeta}^{i,1}) 
		+ &\tilde{Q}_i(\boldsymbol{\zeta}^{i,2})\tilde{\pi}(\boldsymbol{\zeta}^{i,2}) + \tilde{f}_i(\boldsymbol{\zeta}^{i,3})\tilde{\pi}(\boldsymbol{\zeta}^{i,3}) \\&- \big(\tilde{q}_i(\boldsymbol{\zeta}) + \tilde{Q}_i(\boldsymbol{\zeta})  + \tilde{f}_i(\boldsymbol{\zeta})\big)\tilde{\pi}(\boldsymbol{\zeta})\bigg) = 0.
	\end{split}
\end{equation}

Before writing the local divergence condition for the headway process, which is equivalent to \eqref{local_divergence}, we establish notation
\begin{align}
	\tilde{D}_i(\boldsymbol{\zeta}) &= \tilde{q}_i(\boldsymbol{\zeta}^{i,1})\dfrac{\tilde{\pi}(\boldsymbol{\zeta}^{i,1})}{\tilde{\pi}(\boldsymbol{\zeta})} -\tilde{q}_i(\boldsymbol{\zeta}),\\
	\tilde{E}_i(\boldsymbol{\zeta}) &= \tilde{Q}_i(\boldsymbol{\zeta}^{i,2})\dfrac{\tilde{\pi}(\boldsymbol{\zeta}^{i,2})}{\tilde{\pi}(\boldsymbol{\zeta})} -\tilde{Q}_i(\boldsymbol{\zeta}),\\
	\tilde{F}_i(\boldsymbol{\zeta}) &= \tilde{f}_i(\boldsymbol{\zeta}^{i,3})\dfrac{\tilde{\pi}(\boldsymbol{\zeta}^{i,3})}{\tilde{\pi}(\boldsymbol{\zeta})} -\tilde{f}_i(\boldsymbol{\zeta}).
\end{align}
One has
\begin{align}
	&\theta_j^p(\boldsymbol{\zeta}^{i,1}) = \delta_{m_j+\delta_{j,i-1}-\delta_{j,i},p} = \theta_j^{p-\delta_{j,i-1}+\delta_{j,i}}(\boldsymbol{\zeta}) \text{ and } \delta_{s_i^{i,1},\alpha} =\delta_{s_i,3-\alpha},\\
	&\theta_j^p(\boldsymbol{\zeta}^{i,2}) = \delta_{m_j+\delta_{j,i-1}-\delta_{j,i},p} = \theta_j^{p-\delta_{j,i-1}+\delta_{j,i}}(\boldsymbol{\zeta}) \text{ and } \delta_{s_i^{i,2},\alpha} =\delta_{s_i,\alpha},\\
	&\theta_j^p(\boldsymbol{\zeta}^{i,3}) = \theta_j^{p}(\boldsymbol{\zeta}) \text{ and } \delta_{s_i^{i,3},\alpha} =\delta_{s_i,3-\alpha}.
\end{align}
Thus, one gets
\begin{align}
	\dfrac{\tilde{\pi}(\boldsymbol{\zeta}^{i,1})}{\tilde{\pi}(\boldsymbol{\zeta})} & =x^{3-2s_{i}}y^{\theta_{i-1}^{0}-\theta_{i-1}^{1}+\theta_{i}^{0}},\\
	\dfrac{\tilde{\pi}(\boldsymbol{\zeta}^{i,2})}{\tilde{\pi}(\boldsymbol{\zeta})} & =y^{\theta_{i-1}^{0}-\theta_{i-1}^{1}+\theta_{i}^{0}},\\
	\dfrac{\tilde{\pi}(\boldsymbol{\zeta}^{i,3})}{\tilde{\pi}(\boldsymbol{\zeta})} & =x^{3-2s_{i}}.
\end{align}
Similarly to \eqref{local_divergence}, the master equation $\eqref{mas_eq4}$ is satisfied if one requires
\begin{equation}\label{Noether}
	\tilde{D}_i + \tilde{E}_i + \tilde{F}_i  = \tilde{\Phi}_{i-1} - \tilde{\Phi}_{i},
\end{equation}
where $\tilde{\Phi}_i$ is of the form $\tilde{\Phi}_i = (a+ b\theta_{i}^0 + c\theta_{i}^1)(\delta_{s_i,1}+\delta_{s_i,2}) = a+ b\theta_{i}^0 + c\theta_{i}^1$. Note that $\tilde{\Phi}_i$ must conform to this form, as $\tilde{D}_i$, $\tilde{E}_i$, and $\tilde{F}_i$ are contingent upon the state of particle $i$ and variables $\theta_{i-1}^0$, $\theta_{i-1}^1$, $\theta_{i}^0$, $\theta_{i}^1$, all of which belong to the set $\{0,1\}$. By examining all possible cases of \eqref{Noether}, one can initially obtain

\begin{align}
	b & = \dfrac{-q^{\star}-xQ^{\star}}{1+x},\label{value_b}\\
	c & = \dfrac{q^{\star}q^{\star01}+xQ^{\star}Q^{\star01}}{1+x},\label{value_c}
\end{align}
and then one gets the stationary conditions \eqref{cond1}--\eqref{cond2}. However, we shall give clearer proof of the above claim in the next subsection after knowing the results.

\subsection{Proof of Theorem \ref{main_theorem}}
To prove Theorem \ref{main_theorem}, it suffices to demonstrate that \eqref{Noether} holds for all configurations. Note that \eqref{Noether} depends on distance variables $\theta_{i-1}^0, \theta_{i-1}^1, \theta_{i}^0, \theta_{i}^1$, or equivalently, on $m_{i-1}$ and $m_i$. Let us consider the various cases of $m_{i-1}$ and $m_i$ as follows:

\begin{itemize}
	\item  If $m_{i-1}$ and $m_{i}$ are both greater than 1, one has $	\tilde{D}_i  = x^{-1}q^\star\delta_{s_i,2} -q^\star\delta_{s_i,1}, \tilde{E}_i = 0,
	\tilde{F}_i  = xf^\star\delta_{s_i,1} -f^\star\delta_{s_i,2}$. Note that in this case, the right-hand side of \eqref{Noether} is 0. If the state of the particle is 1, meaning $s_i = 1$, then $\tilde{D}_i + \tilde{E}_i + \tilde{F}_i = -q^\star + xf^\star$ which is equal to 0 due to \eqref{cond1}. Similarly, for $s_i = 2$, $\tilde{D}_i + \tilde{E}_i + \tilde{F}_i = x^{-1}q^\star - f^\star$, also equaling to 0 due to \eqref{cond1}. Thus, \eqref{Noether} holds in this case.
	\item If $m_{i-1}>1, m_{i} = 1$, one has $	\tilde{D}_i  = x^{-1}q^\star\delta_{s_i,2} -q^\star\delta_{s_i,1}(1+q^{\star01})$, $ \tilde{E}_i = Q^\star\delta_{s_i,2} -Q^\star\delta_{s_i,2}(1+Q^{\star01})$, $
	\tilde{F}_i  = (x\delta_{s_i,1} -\delta_{s_i,2})f^\star(1+f^{\star01})$. In this case, the right-hand side of \eqref{Noether} is $-c$. If $s_i = 1$, then $\tilde{D}_i + \tilde{E}_i + \tilde{F}_i = -q^\star(1+q^{\star01}) + xf^\star(1+f^{\star01})$.  It can be verified from \eqref{cond1} and \eqref{cond2} that $\tilde{D}_i + \tilde{E}_i + \tilde{F}_i = -c$, where $c$ is defined in \eqref{value_c}. Thus, \eqref{Noether} holds for this case. Similarly, for $s_i = 2$, \eqref{Noether} holds as well.
	\item For the remaining cases ($m_{i-1} > 1, m_{i} = 0$; $m_{i-1} = 1, m_{i} > 1$; $m_{i-1} = 0, m_{i} > 1$; $m_{i-1} = 1, m_{i} = 1$; $m_{i-1} = 1, m_{i} = 0$; $m_{i-1} = 0, m_{i} = 1$; $m_{i-1} = 0, m_{i} = 0$), similar reasoning demonstrates that \eqref{Noether} holds.
\end{itemize}
Thus, the proof is complete.

\section{Partition function and headway distribution}

Just like in previous studies \cite{Belitsky2019-1,Belitsky2019-2}, instead of using measure \eqref{inv_meas_1}, we adopt the grand-canonical ensemble for ease of analysis, as defined by
\begin{equation}\label{grand_can_ensem}
	\tilde{\pi}(\zeta)=\dfrac{1}{Z_{gc}}\prod_{i=1}^{N}x^{-3/2+s_{i}}y^{-\theta_{i}^{0}}z^{m_{i}},
\end{equation}
where $Z_{gc}=(Z_{1}Z_{2})^{N}$ with
\begin{equation}
	Z_{1}=\dfrac{1+(y-1)z}{1-z},\ \ Z_{2}=x^{1/2}+x^{-1/2}.
\end{equation}
Here, the fugacity $z$ serves as a fugacity through which the particle density can be determined.

As one can observe, the partition function $Z_{gc}$ has a fixed form. Consequently, it is necessary to demonstrate the well-defined nature of the measure \eqref{grand_can_ensem}. To achieve this, it is crucial to establish the equivalence between the two measures \eqref{inv_meas_2} and \eqref{grand_can_ensem}, wherein a value of $z$ corresponding to a given particle density $\rho_p$ can always be found, ensuring that the probability of a configuration under both measures is identical. It is worth noting that this task was previously addressed in \cite{Belitsky2019-1, Belitsky2019-2}, given the similarity between the grand-canonical ensemble form \eqref{grand_can_ensem} and the one presented in the aforementioned papers. However, for the benefit of our readers, we rewrite the proof here. On the one hand, one can compute the mean headway using the grand-ensemble \eqref{grand_can_ensem}. Namely, one has
\begin{equation}
	\langle m_{i}\rangle=\dfrac{1}{N}z\dfrac{d}{dz}\ln Z_{gc}=\dfrac{yz}{(1-z)(1+(y-1)z)}.
\end{equation}
Here, $\langle\cdot \rangle$ indicates taking the expectation with respect to the distribution \eqref{grand_can_ensem}. On the other hand, the mean headway can be computed as $\langle m_{i}\rangle =\dfrac{1}{\rho_p}-1$, representing the average number of empty sites on the lattice per particle. Here, $\rho_p$ is the particle density. Thus, the value of $z$ can be determined by solving the following equation
\begin{equation}
	(y-1)z^{2}+z\big(\dfrac{y\rho_p}{1-\rho_p}-y+2\big)-1=0.\label{eq_value_z}
\end{equation}
Note that the quadratic equation above yields two solutions for $z$. However, the chosen value of $z$ must ensure that $Z_{gc}$ remains finite. This condition implies that $0 \leq z < 1$. Considering that $0 < \rho_p \leq 1$, the value of $z$ can be selected as
\begin{equation}
	z(\rho_p,y)=1-\dfrac{1-\sqrt{1-4\rho_p(1-\rho_p)(1-y^{-1})}}{2(1-\rho_p)(1-y^{-1})}\label{value_z}.
\end{equation}
This selection guarantees the condition $0 \leq z < 1$.

When working with the grand-canonical ensemble \eqref{grand_can_ensem} concerning the headway process, deriving the probability distribution for the gap between two particles is easily achieved, defined as
\begin{equation}
	P_h(r) =\tilde{\pi}_{gc}(m_i = r).
\end{equation}
Alternatively, we can express $P_h(r)$ as $\langle \theta_i^r\rangle$, where the random variables $\theta_i^r$ are defined in equation \eqref{newvar}. Therefore, the headway distribution can be expressed as

\begin{equation}\label{headway_distribution}P_h(r) = 
	\begin{cases}
		\dfrac{1-z}{1+(y-1)z} \ \ \text{ for } r = 0, \\
		yP_h(0)z^r \ \ \text{ for } r \geq 1.
	\end{cases}
\end{equation}


\end{document}